\documentclass[longauth]{aa}
\usepackage[varg]{txfonts}
\usepackage{graphicx}
\usepackage{graphics}
\usepackage{amssymb}
\usepackage{amsmath}
\usepackage{url,hyperref}
\usepackage{footmisc}
\usepackage[T1]{fontenc}
\usepackage[utf8]{inputenc}
\newcommand{\HF}[1]{\; \mbox{H}\left[ #1 \right]}  
\newcommand{\stat}{_{\rm stat}}
\newcommand{\sys}{_{\rm sys}}
\newcommand{\fermi}{\textit{Fermi}-LAT}
\newcommand{\source}{\object{3C\,279}}
\newcommand{\hess}{H.E.S.S.}
\newcommand{\zr}{z}
\newcommand{\est}[3]{\left( \frac{#1}{#2} \right)^{#3}}
\newcommand{\p}{^{\prime}}
\newcommand{\as}{^{\ast}}
\newcommand{\E}[1]{\times 10^{#1}}
\newcommand{\g}{\ensuremath{\gamma}}
\makeatletter
\renewcommand*{\@fnsymbol}[1]{\ifcase#1\or*\or$\dagger$\or$\ddagger$\or**\or$\dagger\dagger$\or$\ddagger\ddagger$\fi}
\makeatother
\begin{document}
%
\title{Constraints on the emission region of \source\ during strong flares in 2014 and 2015 through VHE $\gamma$-ray observations with H.E.S.S.}
\author{H.E.S.S. Collaboration
\and H.~Abdalla \inst{\ref{NWU}}
\and R.~Adam \inst{\ref{LLR}}
\and F.~Aharonian \inst{\ref{MPIK},\ref{DIAS},\ref{RAU}}
\and F.~Ait~Benkhali \inst{\ref{MPIK}}
\and E.O.~Ang\"uner \inst{\ref{CPPM}}
\and M.~Arakawa \inst{\ref{Rikkyo}}
\and C.~Arcaro \inst{\ref{NWU}}
\and C.~Armand \inst{\ref{LAPP}}
\and H.~Ashkar \inst{\ref{IRFU}}
\and M.~Backes \inst{\ref{UNAM},\ref{NWU}}
\and V.~Barbosa~Martins \inst{\ref{DESY}}
\and M.~Barnard \inst{\ref{NWU}}
\and Y.~Becherini \inst{\ref{Linnaeus}}
\and D.~Berge \inst{\ref{DESY}}
\and K.~Bernl\"ohr \inst{\ref{MPIK}}
\and R.~Blackwell \inst{\ref{Adelaide}}
\and M.~B\"ottcher \inst{\ref{NWU}}
\and C.~Boisson \inst{\ref{LUTH}}
\and J.~Bolmont \inst{\ref{LPNHE}}
\and S.~Bonnefoy \inst{\ref{DESY}}
\and J.~Bregeon \inst{\ref{LUPM}}
\and M.~Breuhaus \inst{\ref{MPIK}}
\and F.~Brun \inst{\ref{IRFU}}
\and P.~Brun \inst{\ref{IRFU}}
\and M.~Bryan \inst{\ref{GRAPPA}}
\and M.~B\"{u}chele \inst{\ref{ECAP}}
\and T.~Bulik \inst{\ref{UWarsaw}}
\and T.~Bylund \inst{\ref{Linnaeus}}
\and M.~Capasso \inst{\ref{IAAT}}
\and S.~Caroff \inst{\ref{LPNHE}}
\and A.~Carosi \inst{\ref{LAPP}}
\and S.~Casanova \inst{\ref{IFJPAN},\ref{MPIK}}
\and M.~Cerruti \inst{\ref{LPNHE},\ref{CerrutiNowAt}}
\and T.~Chand \inst{\ref{NWU}}
\and S.~Chandra \inst{\ref{NWU}}
\and A.~Chen \inst{\ref{WITS}}
\and S.~Colafrancesco \inst{\ref{WITS}} \protect\footnotemark[2] 
\and M.~Cury{\l}o \inst{\ref{UWarsaw}}
\and I.D.~Davids \inst{\ref{UNAM}}
\and C.~Deil \inst{\ref{MPIK}}
\and J.~Devin \inst{\ref{CENBG}}
\and P.~deWilt \inst{\ref{Adelaide}}
\and L.~Dirson \inst{\ref{HH}}
\and A.~Djannati-Ata\"i \inst{\ref{APC}}
\and A.~Dmytriiev \inst{\ref{LUTH}}
\and A.~Donath \inst{\ref{MPIK}}
\and V.~Doroshenko \inst{\ref{IAAT}}
\and L.O'C.~Drury \inst{\ref{DIAS}}
\and J.~Dyks \inst{\ref{NCAC}}
\and K.~Egberts \inst{\ref{UP}}
\and G.~Emery \inst{\ref{LPNHE}}
\and J.-P.~Ernenwein \inst{\ref{CPPM}}
\and S.~Eschbach \inst{\ref{ECAP}}
\and K.~Feijen \inst{\ref{Adelaide}}
\and S.~Fegan \inst{\ref{LLR}}
\and A.~Fiasson \inst{\ref{LAPP}}
\and G.~Fontaine \inst{\ref{LLR}}
\and S.~Funk \inst{\ref{ECAP}}
\and M.~F\"u{\ss}ling \inst{\ref{DESY}}
\and S.~Gabici \inst{\ref{APC}}
\and Y.A.~Gallant \inst{\ref{LUPM}}
\and F.~Gat{\'e} \inst{\ref{LAPP}}
\and G.~Giavitto \inst{\ref{DESY}}
\and D.~Glawion \inst{\ref{LSW}}
\and J.F.~Glicenstein \inst{\ref{IRFU}}
\and D.~Gottschall \inst{\ref{IAAT}}
\and M.-H.~Grondin \inst{\ref{CENBG}}
\and J.~Hahn \inst{\ref{MPIK}}
\and M.~Haupt \inst{\ref{DESY}}
\and G.~Heinzelmann \inst{\ref{HH}}
\and G.~Henri \inst{\ref{Grenoble}}
\and G.~Hermann \inst{\ref{MPIK}}
\and J.A.~Hinton \inst{\ref{MPIK}}
\and W.~Hofmann \inst{\ref{MPIK}}
\and C.~Hoischen \inst{\ref{UP}}
\and T.~L.~Holch \inst{\ref{HUB}}
\and M.~Holler \inst{\ref{LFUI}}
\and D.~Horns \inst{\ref{HH}}
\and D.~Huber \inst{\ref{LFUI}}
\and H.~Iwasaki \inst{\ref{Rikkyo}}
\and M.~Jamrozy \inst{\ref{UJK}}
\and D.~Jankowsky \inst{\ref{ECAP}}
\and F.~Jankowsky \inst{\ref{LSW}} \protect\footnotemark[1]
\and A.~Jardin-Blicq \inst{\ref{MPIK}}
\and I.~Jung-Richardt \inst{\ref{ECAP}}
\and M.A.~Kastendieck \inst{\ref{HH}}
\and K.~Katarzy{\'n}ski \inst{\ref{NCUT}}
\and M.~Katsuragawa \inst{\ref{KAVLI}}
\and U.~Katz \inst{\ref{ECAP}}
\and D.~Khangulyan \inst{\ref{Rikkyo}}
\and B.~Kh\'elifi \inst{\ref{APC}}
\and J.~King \inst{\ref{LSW}}
\and S.~Klepser \inst{\ref{DESY}}
\and W.~Klu\'{z}niak \inst{\ref{NCAC}}
\and Nu.~Komin \inst{\ref{WITS}}
\and K.~Kosack \inst{\ref{IRFU}}
\and D.~Kostunin \inst{\ref{DESY}} 
\and M.~Kraus \inst{\ref{ECAP}}
\and G.~Lamanna \inst{\ref{LAPP}}
\and J.~Lau \inst{\ref{Adelaide}}
\and A.~Lemi\`ere \inst{\ref{APC}}
\and M.~Lemoine-Goumard \inst{\ref{CENBG}}
\and J.-P.~Lenain \inst{\ref{LPNHE}}
\and E.~Leser \inst{\ref{UP},\ref{DESY}}
\and C.~Levy \inst{\ref{LPNHE}}
\and T.~Lohse \inst{\ref{HUB}}
\and I.~Lypova \inst{\ref{DESY}}
\and J.~Mackey \inst{\ref{DIAS}}
\and J.~Majumdar \inst{\ref{DESY}}
\and D.~Malyshev \inst{\ref{IAAT}}
\and V.~Marandon \inst{\ref{MPIK}}
\and A.~Marcowith \inst{\ref{LUPM}}
\and A.~Mares \inst{\ref{CENBG}}
\and C.~Mariaud \inst{\ref{LLR}}
\and G.~Mart\'i-Devesa \inst{\ref{LFUI}}
\and R.~Marx \inst{\ref{MPIK}}
\and G.~Maurin \inst{\ref{LAPP}}
\and P.J.~Meintjes \inst{\ref{UFS}}
\and A.M.W.~Mitchell \inst{\ref{MPIK},\ref{MitchellNowAt}}
\and R.~Moderski \inst{\ref{NCAC}}
\and M.~Mohamed \inst{\ref{LSW}}
\and L.~Mohrmann \inst{\ref{ECAP}}
\and C.~Moore \inst{\ref{Leicester}}
\and E.~Moulin \inst{\ref{IRFU}}
\and J.~Muller \inst{\ref{LLR}}
\and T.~Murach \inst{\ref{DESY}}
\and S.~Nakashima  \inst{\ref{RIKKEN}}
\and M.~de~Naurois \inst{\ref{LLR}}
\and H.~Ndiyavala  \inst{\ref{NWU}}
\and F.~Niederwanger \inst{\ref{LFUI}}
\and J.~Niemiec \inst{\ref{IFJPAN}}
\and L.~Oakes \inst{\ref{HUB}}
\and P.~O'Brien \inst{\ref{Leicester}}
\and H.~Odaka \inst{\ref{Tokyo}}
\and S.~Ohm \inst{\ref{DESY}}
\and E.~de~Ona~Wilhelmi \inst{\ref{DESY}}
\and M.~Ostrowski \inst{\ref{UJK}}
\and I.~Oya \inst{\ref{DESY}}
\and M.~Panter \inst{\ref{MPIK}}
\and R.D.~Parsons \inst{\ref{MPIK}}
\and C.~Perennes \inst{\ref{LPNHE}}
\and P.-O.~Petrucci \inst{\ref{Grenoble}}
\and B.~Peyaud \inst{\ref{IRFU}}
\and Q.~Piel \inst{\ref{LAPP}}
\and S.~Pita \inst{\ref{APC}}
\and V.~Poireau \inst{\ref{LAPP}}
\and A.~Priyana~Noel \inst{\ref{UJK}}
\and D.A.~Prokhorov \inst{\ref{WITS}}
\and H.~Prokoph \inst{\ref{DESY}}
\and G.~P\"uhlhofer \inst{\ref{IAAT}}
\and M.~Punch \inst{\ref{APC},\ref{Linnaeus}}
\and A.~Quirrenbach \inst{\ref{LSW}}
\and S.~Raab \inst{\ref{ECAP}}
\and R.~Rauth \inst{\ref{LFUI}}
\and A.~Reimer \inst{\ref{LFUI}}
\and O.~Reimer \inst{\ref{LFUI}}
\and Q.~Remy \inst{\ref{LUPM}}
\and M.~Renaud \inst{\ref{LUPM}}
\and F.~Rieger \inst{\ref{MPIK}}
\and L.~Rinchiuso \inst{\ref{IRFU}}
\and C.~Romoli \inst{\ref{MPIK}} \protect\footnotemark[1]
\and G.~Rowell \inst{\ref{Adelaide}}
\and B.~Rudak \inst{\ref{NCAC}}
\and E.~Ruiz-Velasco \inst{\ref{MPIK}}
\and V.~Sahakian \inst{\ref{YPI}}
\and S.~Saito \inst{\ref{Rikkyo}}
\and D.A.~Sanchez \inst{\ref{LAPP}}
\and A.~Santangelo \inst{\ref{IAAT}}
\and M.~Sasaki \inst{\ref{ECAP}}
\and R.~Schlickeiser \inst{\ref{RUB}}
\and F.~Sch\"ussler \inst{\ref{IRFU}}
\and A.~Schulz \inst{\ref{DESY}}
\and H.~Schutte \inst{\ref{NWU}}
\and U.~Schwanke \inst{\ref{HUB}}
\and S.~Schwemmer \inst{\ref{LSW}}
\and M.~Seglar-Arroyo \inst{\ref{IRFU}}
\and M.~Senniappan \inst{\ref{Linnaeus}}
\and A.S.~Seyffert \inst{\ref{NWU}}
\and N.~Shafi \inst{\ref{WITS}}
\and K.~Shiningayamwe \inst{\ref{UNAM}}
\and R.~Simoni \inst{\ref{GRAPPA}}
\and A.~Sinha \inst{\ref{APC}}
\and H.~Sol \inst{\ref{LUTH}}
\and A.~Specovius \inst{\ref{ECAP}}
\and M.~Spir-Jacob \inst{\ref{APC}}
\and {\L.}~Stawarz \inst{\ref{UJK}}
\and R.~Steenkamp \inst{\ref{UNAM}}
\and C.~Stegmann \inst{\ref{UP},\ref{DESY}}
\and C.~Steppa \inst{\ref{UP}}
\and T.~Takahashi  \inst{\ref{KAVLI}}
\and T.~Tavernier \inst{\ref{IRFU}}
\and A.M.~Taylor \inst{\ref{DESY}}
\and R.~Terrier \inst{\ref{APC}}
\and D.~Tiziani \inst{\ref{ECAP}}
\and M.~Tluczykont \inst{\ref{HH}}
\and C.~Trichard \inst{\ref{LLR}}
\and M.~Tsirou \inst{\ref{LUPM}}
\and N.~Tsuji \inst{\ref{Rikkyo}}
\and R.~Tuffs \inst{\ref{MPIK}}
\and Y.~Uchiyama \inst{\ref{Rikkyo}}
\and D.J.~van~der~Walt \inst{\ref{NWU}}
\and C.~van~Eldik \inst{\ref{ECAP}}
\and C.~van~Rensburg \inst{\ref{NWU}}
\and B.~van~Soelen \inst{\ref{UFS}}
\and G.~Vasileiadis \inst{\ref{LUPM}}
\and J.~Veh \inst{\ref{ECAP}}
\and C.~Venter \inst{\ref{NWU}}
\and P.~Vincent \inst{\ref{LPNHE}}
\and J.~Vink \inst{\ref{GRAPPA}}
\and F.~Voisin \inst{\ref{Adelaide}}
\and H.J.~V\"olk \inst{\ref{MPIK}}
\and T.~Vuillaume \inst{\ref{LAPP}}
\and Z.~Wadiasingh \inst{\ref{NWU}}
\and S.J.~Wagner \inst{\ref{LSW}}
\and R.~White \inst{\ref{MPIK}}
\and A.~Wierzcholska \inst{\ref{IFJPAN},\ref{LSW}} \protect\footnotemark[1]
\and R.~Yang \inst{\ref{MPIK}}
\and H.~Yoneda \inst{\ref{KAVLI}}
\and M.~Zacharias \inst{\ref{NWU}} \protect\footnotemark[1]
\and R.~Zanin \inst{\ref{MPIK}}
\and A.A.~Zdziarski \inst{\ref{NCAC}}
\and A.~Zech \inst{\ref{LUTH}}
\and A.~Ziegler \inst{\ref{ECAP}}
\and J.~Zorn \inst{\ref{MPIK}}
\and N.~\.Zywucka \inst{\ref{NWU}}
\and \newline M.~Meyer \inst{\ref{MeyerNowAt}}
}
\institute{
Centre for Space Research, North-West University, Potchefstroom 2520, South Africa \label{NWU} \and 
Universit\"at Hamburg, Institut f\"ur Experimentalphysik, Luruper Chaussee 149, D 22761 Hamburg, Germany \label{HH} \and 
Max-Planck-Institut f\"ur Kernphysik, P.O. Box 103980, D 69029 Heidelberg, Germany \label{MPIK} \and 
Dublin Institute for Advanced Studies, 31 Fitzwilliam Place, Dublin 2, Ireland \label{DIAS} \and 
High Energy Astrophysics Laboratory, RAU,  123 Hovsep Emin St  Yerevan 0051, Armenia \label{RAU} \and
Yerevan Physics Institute, 2 Alikhanian Brothers St., 375036 Yerevan, Armenia \label{YPI} \and
Institut f\"ur Physik, Humboldt-Universit\"at zu Berlin, Newtonstr. 15, D 12489 Berlin, Germany \label{HUB} \and
University of Namibia, Department of Physics, Private Bag 13301, Windhoek, Namibia, 12010 \label{UNAM} \and
GRAPPA, Anton Pannekoek Institute for Astronomy, University of Amsterdam,  Science Park 904, 1098 XH Amsterdam, The Netherlands \label{GRAPPA} \and
Department of Physics and Electrical Engineering, Linnaeus University,  351 95 V\"axj\"o, Sweden \label{Linnaeus} \and
Institut f\"ur Theoretische Physik, Lehrstuhl IV: Weltraum und Astrophysik, Ruhr-Universit\"at Bochum, D 44780 Bochum, Germany \label{RUB} \and
Institut f\"ur Astro- und Teilchenphysik, Leopold-Franzens-Universit\"at Innsbruck, A-6020 Innsbruck, Austria \label{LFUI} \and
School of Physical Sciences, University of Adelaide, Adelaide 5005, Australia \label{Adelaide} \and
LUTH, Observatoire de Paris, PSL Research University, CNRS, Universit\'e Paris Diderot, 5 Place Jules Janssen, 92190 Meudon, France \label{LUTH} \and
Sorbonne Universit\'e, Universit\'e Paris Diderot, Sorbonne Paris Cit\'e, CNRS/IN2P3, Laboratoire de Physique Nucl\'eaire et de Hautes Energies, LPNHE, 4 Place Jussieu, F-75252 Paris, France \label{LPNHE} \and
Laboratoire Univers et Particules de Montpellier, Universit\'e Montpellier, CNRS/IN2P3,  CC 72, Place Eug\`ene Bataillon, F-34095 Montpellier Cedex 5, France \label{LUPM} \and
IRFU, CEA, Universit\'e Paris-Saclay, F-91191 Gif-sur-Yvette, France \label{IRFU} \and
Astronomical Observatory, The University of Warsaw, Al. Ujazdowskie 4, 00-478 Warsaw, Poland \label{UWarsaw} \and
Aix Marseille Universit\'e, CNRS/IN2P3, CPPM, Marseille, France \label{CPPM} \and
Instytut Fizyki J\c{a}drowej PAN, ul. Radzikowskiego 152, 31-342 Krak{\'o}w, Poland \label{IFJPAN} \and
School of Physics, University of the Witwatersrand, 1 Jan Smuts Avenue, Braamfontein, Johannesburg, 2050 South Africa \label{WITS} \and
Laboratoire d'Annecy de Physique des Particules, Univ. Grenoble Alpes, Univ. Savoie Mont Blanc, CNRS, LAPP, 74000 Annecy, France \label{LAPP} \and
Landessternwarte, Universit\"at Heidelberg, K\"onigstuhl, D 69117 Heidelberg, Germany \label{LSW} \and
Universit\'e Bordeaux, CNRS/IN2P3, Centre d'\'Etudes Nucl\'eaires de Bordeaux Gradignan, 33175 Gradignan, France \label{CENBG} \and
Oskar Klein Centre, Department of Physics, Stockholm University, Albanova University Center, SE-10691 Stockholm, Sweden \label{OKC} \and
Institut f\"ur Astronomie und Astrophysik, Universit\"at T\"ubingen, Sand 1, D 72076 T\"ubingen, Germany \label{IAAT} \and
Laboratoire Leprince-Ringuet, Ã‰cole Polytechnique, UMR 7638, CNRS/IN2P3, Institut Polytechnique de Paris, F-91128 Palaiseau, France \label{LLR} \and
APC, AstroParticule et Cosmologie, Universit\'{e} Paris Diderot, CNRS/IN2P3, CEA/Irfu, Observatoire de Paris, Sorbonne Paris Cit\'{e}, 10, rue Alice Domon et L\'{e}onie Duquet, 75205 Paris Cedex 13, France \label{APC} \and
Univ. Grenoble Alpes, CNRS, IPAG, F-38000 Grenoble, France \label{Grenoble} \and
Department of Physics and Astronomy, The University of Leicester, University Road, Leicester, LE1 7RH, United Kingdom \label{Leicester} \and
Nicolaus Copernicus Astronomical Center, Polish Academy of Sciences, ul. Bartycka 18, 00-716 Warsaw, Poland \label{NCAC} \and
Institut f\"ur Physik und Astronomie, Universit\"at Potsdam,  Karl-Liebknecht-Strasse 24/25, D 14476 Potsdam, Germany \label{UP} \and
Friedrich-Alexander-Universit\"at Erlangen-N\"urnberg, Erlangen Centre for Astroparticle Physics, Erwin-Rommel-Str. 1, D 91058 Erlangen, Germany \label{ECAP} \and
DESY, D-15738 Zeuthen, Germany \label{DESY} \and
Obserwatorium Astronomiczne, Uniwersytet Jagiello{\'n}ski, ul. Orla 171, 30-244 Krak{\'o}w, Poland \label{UJK} \and
Centre for Astronomy, Faculty of Physics, Astronomy and Informatics, Nicolaus Copernicus University,  Grudziadzka 5, 87-100 Torun, Poland \label{NCUT} \and
Department of Physics, University of the Free State,  PO Box 339, Bloemfontein 9300, South Africa \label{UFS} \and
Department of Physics, Rikkyo University, 3-34-1 Nishi-Ikebukuro, Toshima-ku, Tokyo 171-8501, Japan \label{Rikkyo} \and
Kavli Institute for the Physics and Mathematics of the Universe (WPI), The University of Tokyo Institutes for Advanced Study (UTIAS), The University of Tokyo, 5-1-5 Kashiwa-no-Ha, Kashiwa City, Chiba, 277-8583, Japan \label{KAVLI} \and
Department of Physics, The University of Tokyo, 7-3-1 Hongo, Bunkyo-ku, Tokyo 113-0033, Japan \label{Tokyo} \and
RIKEN, 2-1 Hirosawa, Wako, Saitama 351-0198, Japan \label{RIKKEN} \and
Now at Physik Institut, Universit\"at Z\"urich, Winterthurerstrasse 190, CH-8057 Z\"urich, Switzerland \label{MitchellNowAt} \and
Now at Institut de Ci\`{e}ncies del Cosmos (ICC UB), Universitat de Barcelona (IEEC-UB), Mart\'{i} Franqu\`es 1, E08028 Barcelona, Spain \label{CerrutiNowAt} \and 
W. W. Hansen Experimental Physics Laboratory, Kavli Institute for Particle Astrophysics and Cosmology, Department of Physics and SLAC National Accelerator Laboratory, Stanford University, Stanford, CA 94305, USA \label{MeyerNowAt}
}
\offprints{H.E.S.S.~collaboration,
\protect\\\email{\href{mailto:contact.hess@hess-experiment.eu}{contact.hess@hess-experiment.eu}};
\protect\\\protect\footnotemark[1] Corresponding authors
\protect\\\protect\footnotemark[2] Deceased
}
%
%
\date{Received ??? / accepted ??? }
%
%
\abstract{The flat spectrum radio quasar \source\ is known to exhibit pronounced variability in the high-energy ($100\,$MeV$<E<100\,$GeV) $\gamma$-ray band, which is continuously monitored with \fermi. During two periods of high activity in April 2014 and June 2015 Target-of-Opportunity observations were undertaken with \hess\ in the very-high-energy (VHE, $E>100\,$GeV) $\gamma$-ray domain. While the observation in 2014 provides an upper limit, the observation in 2015 results in a signal with $8.7\,\sigma$ significance above an energy threshold of $66\,$GeV. 
No VHE variability has been detected during the 2015 observations. The VHE photon spectrum is soft and described by a power-law index of $4.2\pm 0.3$.
The \hess\ data along with a detailed and contemporaneous multiwavelength data set provide constraints on the physical parameters of the emission region. The minimum distance of the emission region from the central black hole is estimated using two plausible geometries of the broad-line region and three potential intrinsic spectra. The emission region is confidently placed at $r\gtrsim 1.7\times10^{17}\,$cm from the black hole, i.e., beyond the assumed distance of the broad-line region. 
Time-dependent leptonic and lepto-hadronic one-zone models are used to describe the evolution of the 2015 flare. Neither model can fully reproduce the observations, despite testing various parameter sets.
Furthermore, the \hess\ data are used to derive constraints on Lorentz invariance violation given the large redshift of \source.
}
\titlerunning{Constraints on \source\ through H.E.S.S. observations}
\authorrunning{H.E.S.S. Collaboration et al.}
\maketitle
%
\makeatletter
\renewcommand*{\@fnsymbol}[1]{\ifcase#1\@arabic{#1}\fi}
\makeatother
%
\section{Introduction} \label{sec:intro}
\source\ \citep[redshift $\zr=0.536$,][$\mbox{RA}_{\rm J2000}=12^{\rm h}56^{\rm m}11.1^{\rm s}$, $\mbox{DEC}_{\rm J2000}=-05^{\rm d}47^{\rm m}22^{\rm s}$]{br65,msdcm96} belongs to the class of flat spectrum radio quasars (FSRQs) that are characterized by strong variability in all energy bands from radio to $\gamma$-rays, and broad emission lines (equivalent width $>5\,$\AA) in the optical spectrum signifying the existence of a broad-line region (BLR). FSRQs belong to the blazar class of active galactic nuclei, and their jets are closely aligned with the line of sight \citep{br74} resulting in strongly Doppler-boosted emission. Spectral energy distributions (SEDs) of FSRQs exhibit two broad, non-thermal components. The low-energy component peaks in the infrared and is attributed to electron synchrotron emission. In leptonic scenarios, the high-energy component, which peaks below the GeV regime, is attributed to inverse Compton (IC) emission of the same electrons scattering off ambient soft photon fields. Such soft photon fields can be the synchrotron emission (synchrotron-self Compton, SSC), photons from the accretion disk (IC/Disk), the broad-line region (IC/BLR) or the infrared emission of the dusty torus (IC/DT). In lepto-hadronic models, the high energy spectral component is attributed to processes involving highly relativistic protons, such as proton synchrotron, or secondary emission from photo-meson production. The latter includes synchrotron emission from charged pions, muons, and the resulting secondary electrons and positrons. For a review of these processes see, e.g., \cite{b07}.

While FSRQs are bright in the high-energy (HE, $100\,$MeV$<E<100\,$GeV) $\gamma$-ray domain, they are much fainter at very high-energy (VHE, $E>100\,$GeV) $\gamma$-rays for a number of reasons. Firstly, the low peak energy around the lower end of the HE $\gamma$-ray domain might indicate a low maximum particle Lorentz factor, implying emission well below the VHE regime. Secondly, if the $\gamma$-rays are produced within $\sim 0.1\,$pc from the central supermassive black hole, any VHE emission would be strongly attenuated by the BLR photon field. Observations of VHE emission will therefore allow one to significantly constrain the minimum distance of the emission region from the black hole as the intrinsic absorption by the BLR cannot be too severe. Thirdly, FSRQs are found at rather large cosmological redshifts, with the closest VHE-detected FSRQ at $\zr=0.189$ \citep[PKS\,0736+017,][]{cea17}. Hence, attenuation of VHE $\gamma$-rays by the extragalactic background light (EBL) will also reduce the detectable \g-ray flux. 

\source\ was detected at VHE $\gamma$-rays with MAGIC in 2006 \citep{Mea08} and 2007 \citep{aMea11} during bright optical flares. However, it was not detected at VHE $\gamma$-rays since then \citep{Hea14,aMea14,aVea16}.
In the HE $\gamma$-ray regime, \source\ has been detected with both EGRET \citep{hea99} and \fermi\ \citep{aFea15}. Due to the ongoing monitoring of \fermi, several flares of \source\ have been observed in the last years, of which a few have been followed up with Cherenkov experiments. 

In April 2014 and June 2015, \source\ exhibited strong outbursts in the HE $\gamma$-ray band with integrated fluxes exceeding $10^{-5}\,$ph$\,$cm$^{-2}$s$^{-1}$ on time scales of a few hours \citep{Hea15,p15}. Both flares were observed with \fermi\ in pointing mode, i.e. instead of the usual survey mode, the satellite was pointed towards \source\ to increase the exposure. In the 2015 event, this resulted in the detection of very fast variability on the order of a few minutes \citep{aFea16} on top of the longer-term (several hours) evolution of the event. Both of these events have been followed up with \hess, and the results are reported here. While there is no detection in VHE $\gamma$-rays in 2014, the 2015 observation has resulted in a significant detection.

This paper is organized as follows: Section \ref{sec:hess} describes the analysis of the \hess\ observations of both flares. Given the \hess\ detection in 2015, the analysis of a multiwavelength data set of that event is presented in section \ref{sec:mwl}. Sections \ref{sec:2014} and \ref{sec:2015} are devoted to a discussion and interpretation of both events based on various models, with an emphasis placed on the 2015 event. Limits on Lorentz Invariance Violations (LIV) are derived in section \ref{sec:liv}. The results are summarized in section \ref{sec:sumcon}.

Throughout the paper a $\Lambda$CDM cosmology is used 
with $H_0 = 69.6\,$km$\,$s$^{-1}\,$Mpc$^{-1}$, $\Omega_M = 0.286$, and $\Omega_{\lambda} = 0.714.$ \citep[e.g.,][]{blwh14}. The 
resulting luminosity distance of \source\ is $d_L = 3.11\,$Gpc. 
%
%
\section{H.E.S.S. data analysis} \label{sec:hess}
\begin{figure*}[t]
\centering 
\includegraphics[width=1.00\textwidth]{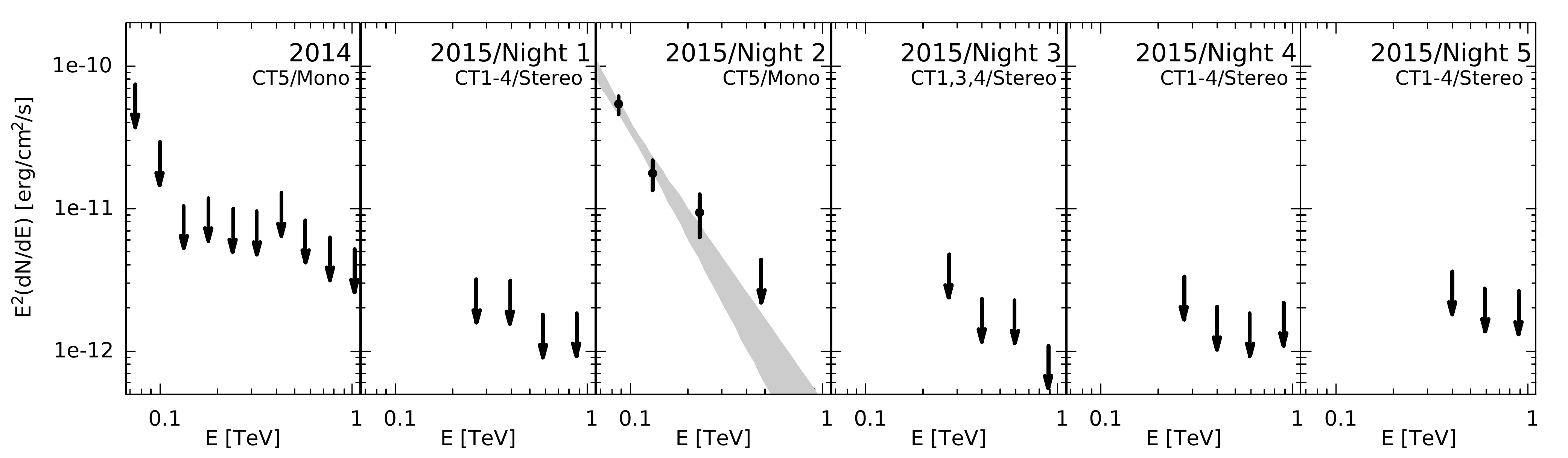}
\caption{Observed \hess\ photon spectra for the six data sets as labeled. Arrows mark upper limits ($99\%$ confidence level). The gray butterfly is the $1\,\sigma$ statistical uncertainty band of the 2015/Night 2 data set. Error bars are statistical only. The second label gives the telescope participation and the analysis used.}
\label{fig:hess1415}
\end{figure*} 
\hess\ is located in the Khomas Highland in Namibia at about 1800\,m above sea level. It is an array of five Imaging Atmospheric Cherenkov Telescopes, with four telescopes (CT\,1-4) with 107\,m$^2$ mirror area arranged in a square of 120\,m side length and one telescope (CT\,5) with 614\,m$^2$ mirror area in the center of the array. Observations are carried out in individual runs of typically $28\,$min duration. For point sources, such as \source, the array observes in wobble mode, i.e. with alternating offsets to the source in right ascension and declination between runs for improved background subtraction. While the array operates in stereo mode -- all telescopes point at the same sky coordinate -- the analysis can be performed for different array layouts depending on the demands of the observed source. A stereo analysis requires that Cherenkov emission has been detected by at least two telescopes, while a mono analysis considers photons detected by CT\,5. A mono analysis with CT\,5 typically provides a lower energy threshold compared to analyses including CT\,1-4 owing to the larger mirror area. The main analysis is performed using the \textsc{Model} analysis chain \citep{dnr09,hea15}. It is cross-checked with an independent calibration chain and the analysis software \textsc{ImPACT} \citep{ph14,pea15}.

In 2014, \hess\ observed \source\ with the full array over three consecutive nights between the 2nd and 4th of April (MJD~$56749$--$56751$) . A mono analysis has been conducted with \verb|very loose| cuts\footnote{The cuts refer to parameter settings for the air shower reconstruction.} \citep{Hea17} resulting in an energy threshold of $66\,$GeV. Seven observation runs passed the quality selection \citep{aHea06}, resulting in $2.6\,$h of acceptance-corrected observation time, and yielding a $3.6\,\sigma$ significance following \cite{lm83}. Differential upper limits ($99\%$ confidence level) have been derived following \cite{fc98} assuming a photon index of 4. The index has been motivated by the detection spectrum of \cite{Mea08}. The upper limits are shown in Fig.~\ref{fig:hess1415}.

Observations in 2015 were conducted in five nights between the 15th and 21st June (MJD~$57188$--$57194$) with changing array configurations. 
During the first night, June 15th (MJD~$57188.7$--$57188.9$, ``Night 1''), CT\,5 was unavailable, and a stereo analysis with \verb|loose| cuts\footnote{Despite the different nomenclature, both mono and stereo analysis cuts imply the lowest possible energy threshold for the respective analyses.} \citep{aHea06} has been conducted on events recorded by CT\,1-4 yielding an energy threshold of $216\,$GeV. Quality selection has resulted in 6 observation runs for the analysis with $2.2\,$h of acceptance corrected observation time and a significance of $1.5\,\sigma$. As for 2014, differential upper limits have been computed with a photon index of 4, c.f. Fig.~\ref{fig:hess1415}. Additionally, an integrated upper limit above $200\,$GeV has been computed, which is shown in the lightcurve in Fig.~\ref{fig:lc-data}(a).

During the second night of observations, June 16th (MJD $57189.7$--$57189.9$, ``Night 2''), CT\,5 was available, and a mono analysis has been conducted with \verb|very loose| cuts and an energy threshold of $66\,$GeV. Quality selection has led to 7 observation runs for the analysis with $2.2\,$h of acceptance corrected observation time, resulting in a detection with $8.7\,\sigma$ significance. 
The spectrum has been modeled assuming a power-law of the form

\begin{align}
 \frac{dN}{dE} = N_0 \left(\frac{E}{E_0} \right)^{-\Gamma} \label{eq:pow}
\end{align}
with normalization $N_0 = (2.5\pm 0.2\stat \pm 0.5\sys)\times 10^{-9}\,$cm$^{-2}$s$^{-1}$TeV$^{-1}$, photon index $\Gamma = 4.2\pm 0.3\stat \pm 0.2\sys$, and decorrelation energy $E_0 = 98\,$GeV; see also Tab.~\ref{tab:fitres}. The systematic errors have been derived following \cite{Hea17}. The spectrum is shown as the gray butterfly ($1\,\sigma$ statistical uncertainty band), points ($>2\,\sigma$ significance level) and arrows ($99\%$ confidence upper limits) in Fig.~\ref{fig:hess1415}. There is no indication for curvature as the goodness-of-fit probability of the power-law spectrum is $p=0.82$. In the following, \hess\ data points that have been corrected for EBL absorption using the EBL model of \cite{frv08}, are used.

The average flux above an energy threshold\footnote{The threshold of $200\,$GeV has been chosen for comparison with the upper limits of the other nights.} of $200\,$GeV equals $(7.6\pm 0.7\stat \pm 1.5\sys)\times 10^{-12}\,$cm$^{-2}$s$^{-1}$, and is shown in Fig.~\ref{fig:lc-data}(a).
A zoom into Night 2 is shown in Fig.~\ref{fig:lc-data-zoom}(a) using run-wise time bins. In order to be comparable to the results of MAGIC in 2006 and 2007 \citep{Mea08,aMea11}, here the lightcurve is derived above an energy threshold of $100\,$GeV. The average flux is $(6.5\pm 0.6\stat \pm 1.3\sys)\times 10^{-11}\,$cm$^{-2}$s$^{-1}$, which is a factor $\sim 10$ less than the flux during the MAGIC detection in 2006 \citep{Mea08}. There is no indication for statistically significant variations in this lightcurve, as a constant flux has a probability of $p=0.39$ ($\chi^2/{\rm ndf}=7.6/6$). 

Observations on June 17th (MJD~$57190.7344$--$57190.8569$, ``Night 3'') were conducted using only CT\,1, 3 and 4. Six runs passed the quality selection, and a stereo analysis with \verb|loose| cuts resulted in a significance of $-0.6\,\sigma$ in $2.3\,$hrs of acceptance corrected observation time. The differential upper limit spectrum (photon index 4) is shown in Fig.~\ref{fig:hess1415}, while the integrated upper limit above an energy threshold of $200\,$GeV is shown in Fig.~\ref{fig:lc-data}(a).

On June 18th (MJD~$57191.7819$--$57191.9193$, ``Night 4'') all five telescopes participated in the observations. However, only 2 of the 5 conducted runs passed the CT\,5 quality selection, which is why a stereo analysis with \verb|loose| cuts has been done on all 5 runs with only the small telescopes. The analysis resulted in a significance of $-2.0\,\sigma$ in $1.7\,$hrs of acceptance corrected observation time. The differential upper limit spectrum is shown in Fig.~\ref{fig:hess1415} and was computed with a photon index of 4, while the integrated upper limit above an energy threshold of $200\,$GeV is given in Fig.~\ref{fig:lc-data}(a).

Two more runs were taken on June 20th (MJD~$57193.8339$--$57193.8740$, ``Night 5'') with all 5 telescopes. However, as in Night 4, the data recorded with CT\,5 did not pass the quality selection. Hence again a stereo analysis with \verb|loose| cuts has been performed on the data recorded with the small telescopes. Due to moon constraints the observations started relatively late, resulting in elevations of less than $52^{\circ}$. This explains the high energy threshold of more than $400\,$GeV in this night. The significance is $-0.3\,\sigma$ in $0.7\,$hrs of acceptance corrected observation time. As before, the differential upper limit spectrum (photon index 4) is shown in Fig.~\ref{fig:hess1415}, while the integrated upper limit above an energy threshold\footnote{This involves an extrapolation to this energy threshold, which is necessary to be comparable with the other nights.} of $200\,$GeV is shown in Fig.~\ref{fig:lc-data}(a).

While the lightcurve shown in Fig.~\ref{fig:lc-data}(a) may be suggestive of variability, the upper limits and the flux point have been achieved with different array configurations. An analysis of Night 2 using only the data from CT\,1-4 results in no detection with an integrated upper limit comparable to the other nights. As the multiwavelength flare subsided after Night 2, and no further detections were achieved with H.E.S.S. after that night, the following discussion will focus on Nights 1 and 2 only.
%
%
\section{Multiwavelength observations of the 2015 flare} \label{sec:mwl}
\begin{figure}[t]
\centering
\includegraphics[width=0.48\textwidth]{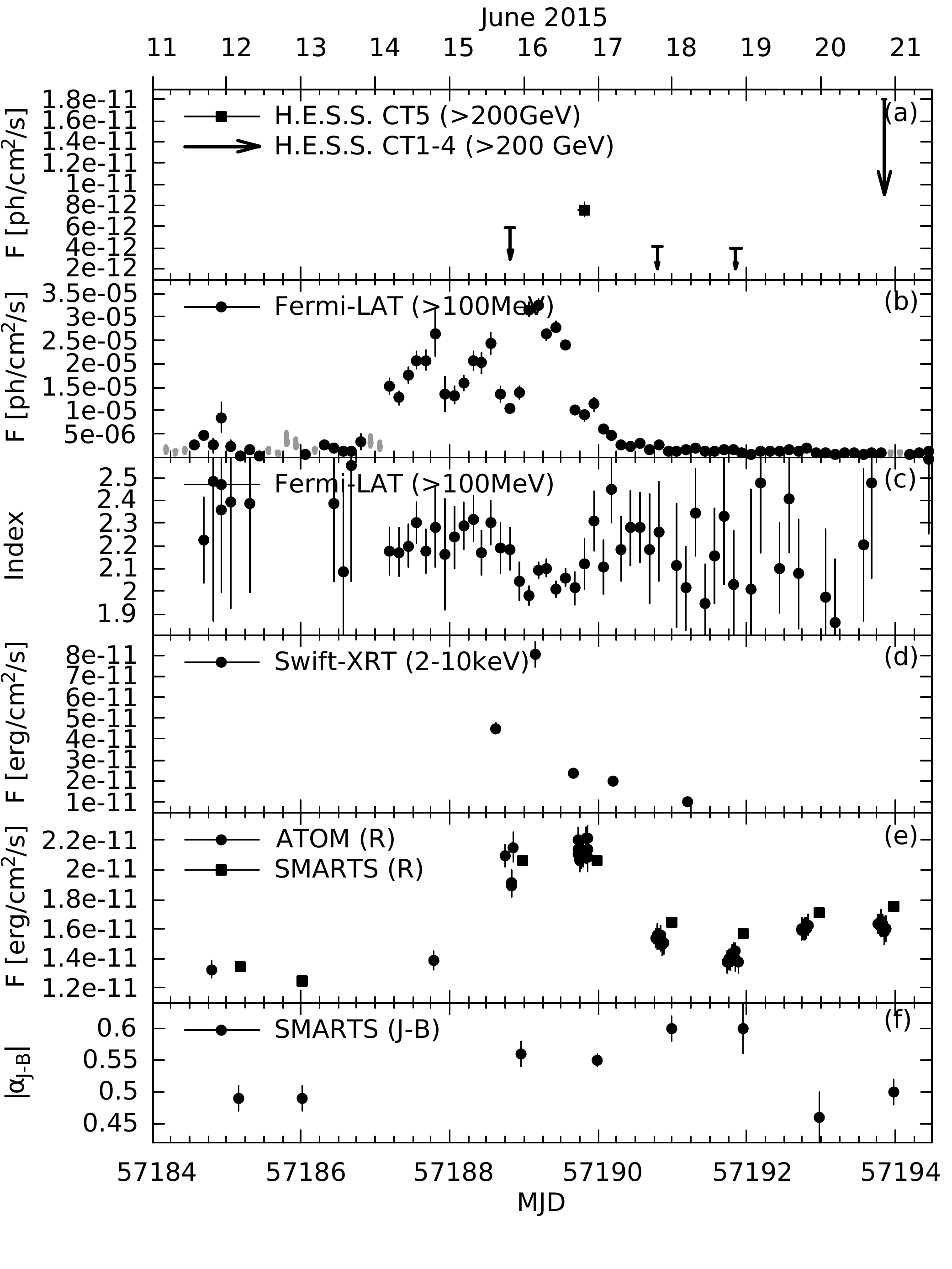}
\caption{Observed multiwavelength lightcurves. 
{\bf (a)} \hess\ lightcurve derived above an energy threshold of $200\,$GeV in night-wise time bins with array configuration as indicated. Arrows mark upper limits (99\% confidence level).
{\bf (b)} \fermi\ lightcurve integrated above $100\,$MeV in $3\,$h bins. Gray arrows mark upper limits (95\% confidence level).
{\bf (c)} HE $\gamma$-ray photon index measured with \fermi\ in $3\,$h bins. 
{\bf (d)} Swift-XRT lightcurve integrated between $2$ and $10\,$keV for individual pointings. 
{\bf (e)} Optical R band lightcurve from ATOM and SMARTS for individual pointings. 
{\bf (f)} Spectral index between the J and B band using SMARTS observations for individual pointings.
In all panels, only statistical error bars are shown.}
\label{fig:lc-data}
\end{figure} 
\begin{figure}[t]
\centering
\includegraphics[width=0.48\textwidth]{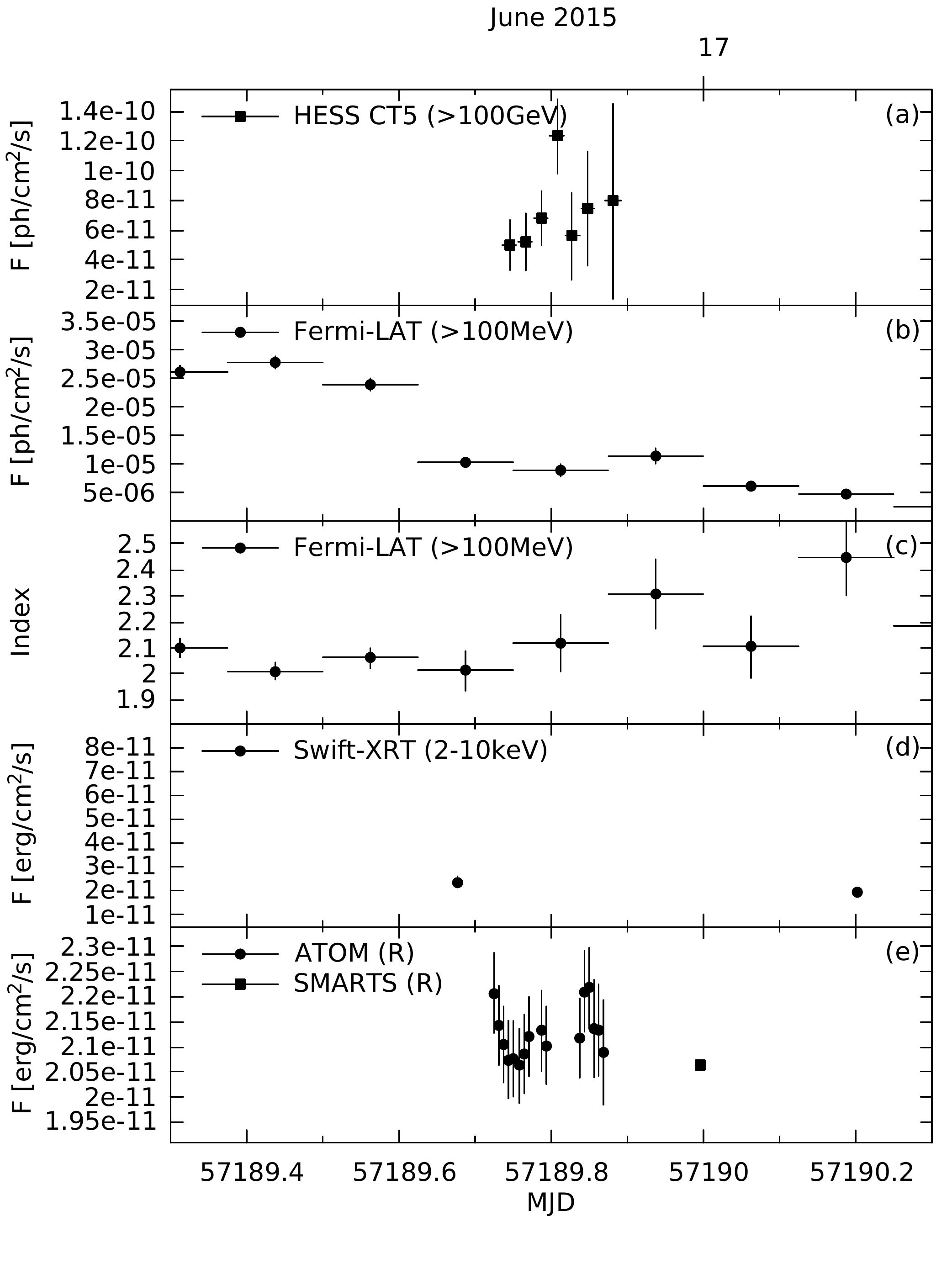}
\caption{Observed multiwavelength lightcurves zoomed in on Night 2. 
{\bf (a)} \hess\ lightcurve derived above an energy threshold of $100\,$GeV in run-wise time bins. 
{\bf (b)} \fermi\ lightcurve integrated above $100\,$MeV in $3\,$h bins. 
{\bf (c)} HE $\gamma$-ray photon index measured with \fermi\ in $3\,$h bins. 
{\bf (d)} Swift-XRT lightcurve integrated between $2$ and $10\,$keV for individual pointings. 
{\bf (e)} Optical R band lightcurve from ATOM and SMARTS for individual pointings. 
In all panels, only statistical error bars are shown, while horizontal bars mark the observation time.}
\label{fig:lc-data-zoom}
\end{figure} 
\begin{table*}[t]
\footnotesize
\caption{Power-law fit of the \hess ($E_0=98$ GeV), \fermi ($E_0=342$ MeV), and \textit{Swift}-XRT ($E_0=1$ keV) observed spectra for the considered time frames. The MJD values give the integration time for the \fermi\ spectra, and the other spectra are chosen to be as contemporaneous as possible. Only statistical errors are given.}
\begin{tabular}{lccccccc}
 \multicolumn{2}{c}{Time frame} 	& \multicolumn{2}{c}{\hess}  						& \multicolumn{2}{c}{\fermi}				 	& \multicolumn{2}{c}{\textit{Swift} XRT} \\
		& MJD			& $N_0$ [ph/cm$^2$/s/TeV]			& $\Gamma_{\rm \hess}$	& $N_0$ [ph/cm$^2$/s/GeV]			& $\Gamma_{\rm LAT}$& $N_0$ [ph/cm$^2$/s/keV]& $\Gamma_{\rm XRT} $ \\
\hline
Preflare	& 57184.0 - 57187.0	& - 						& - 			& $\left(1.1\pm0.1\right)\times 10^{-6}$  	& $2.3\pm 0.1$	& - & - \\
Night 1		& 57188.756 - 57188.880	& \multicolumn{2}{c}{upper limit}		  			& $\left(9.2\pm0.9\right)\times 10^{-6}$ 	& $2.2\pm0.1$	& $\left( 5.9\pm0.3\right)\times 10^{-3}$ & $1.30 \pm 0.05$  \\
Maximum 	& 57189.125 - 57189.250	& -						& - 			& $\left(27\pm1\right)\times 10^{-6}$    	& $2.09\pm0.04$	& $\left( 8.3\pm0.4\right) \times 10^{-3}$ & $1.16 \pm 0.06$ \\
Night 2 	& 57189.734 - 57189.888	& $\left(2.5\pm0.2\right)\times 10^{-9}$	& $4.2\pm0.3$		& $\left(7.7\pm0.8\right)\times 10^{-6}$	& $2.1 \pm 0.1$	& $\left( 3.8\pm0.2\right)\times 10^{-3}$ & $1.43 \pm 0.07$ 
\end{tabular}
\label{tab:fitres}
\end{table*}
In Figs.~\ref{fig:lc-data} and \ref{fig:lc-data-zoom} lightcurves at different wavelengths of the 2015 flare are shown. The analyses are presented below.
\subsection{HE $\gamma$-ray data} \label{sec:fermi}
For the HE band, data taken with the the Large Area Telescope \citep[][LAT]{2009ApJ...697.1071A} on-board the \textit{Fermi} satellite have been analyzed.
The \fermi\ analysis has been carried out using the Science Tool version 10.0.5 and Instrument Response Functions (IRFs) \texttt{P8R2\_SOURCE\_V6}. Data have been analyzed first on a 28 day interval, from MJD~57174 to MJD~57202 using a Binned Analysis method \citep{mea96} on a square region of interest of 30 degree side length and an energy range going from $100\,$MeV to $300\,$GeV. Nearby sources have been modeled using the 3FGL catalog \citep{aFea15} up to a radial distance from the central source of 25 degrees. The spectral parameters of these background sources are kept free if they are within a circle of 5 degrees from the position of \source. In the annulus with angular distances between 5 and 15 degrees only the flux normalization is left free to vary. According to the recommendations of the \fermi\ collaboration, the background models {\tt iso\_P8R2\_SOURCE\_V6\_v06.txt} (isotropic) and {\tt gll\_iem\_v06.fit} (galactic)\footnote{\url{https://fermi.gsfc.nasa.gov/ssc/data/access/lat/BackgroundModels.html}} are used with their normalization fitted to the data.

The lightcurve and spectra for \source\ are obtained by fixing all the background sources in the best fit model obtained from the 28-day time interval, leaving only the spectral parameters for \source\ free to vary. Due to the very high level of photon counts available with \fermi\ for this event, it is possible firstly to perform a detailed 3-hour binned lightcurve of the source near the peak of the emission shown in Figs.~\ref{fig:lc-data}(b) and \ref{fig:lc-data-zoom}(b) along with the photon index in Figs.~\ref{fig:lc-data}(c) and \ref{fig:lc-data-zoom}(c), and secondly to compute the HE $\gamma$-ray spectrum in time intervals strictly simultaneous with the first and second night of the \hess\ observations. In order to create a self-consistent model of the evolution of the flare (see section \ref{sec:2015mod}) two more spectra are produced, namely for the ``Preflare'' time frame and the ``Maximum'' of the \fermi\ lightcurve between Night 1 and Night 2. The precise integration times are given in Tab.~\ref{tab:fitres}. For the calculation of the \fermi\ SED points, a likelihood fit has been performed in the designated energy range, with all free parameters fixed to the best power-law fit values except the normalization of \source. As for lightcurves, a flux point has been computed in case the significance in the bin is above $3\sigma$, a 95\% upper limit has been calculated otherwise, assuming the best-fit power-law photon index over the entire energy range.

In the 3FGL catalogue the HE spectrum is better described by a log-parabola function of the form 

\begin{align}
\frac{dN}{dE}=N_0\left(\frac{E}{E_0}\right)^{-\left(\Gamma+\beta\log\frac{E}{E_0}\right)} \label{eq:curvedspec}
\end{align}
with the curvature parameter $\beta$. In the short time intervals of the observations considered here, only for the Maximum time frame a curved spectrum is preferred on a $4\sigma$ significance level over a power-law. The fit parameters are as follows: $N_0 = 31 \pm 2\times 10^{-5}$ ph/cm$^2$/s/GeV, $\Gamma_{LAT} = 1.96\pm0.05$, and $\beta_{LAT} = 0.12\pm0.03$ at an energy scale $E_0=0.342$ GeV. The best fit spectral values using a power-law, Eq.~(\ref{eq:pow}), are reported in Tab.~\ref{tab:fitres}.
%
%
\subsection{X-ray data}
\begin{table}[t]
\footnotesize
\caption{\textit{Swift}-XRT observations of \source\ used for the time frames defined in Tab.~\ref{tab:fitres}. The columns give the time frame, the Observation ID, the start time and the duration of the observation. The last column gives the UVOT filter.}
\begin{tabular}{lcccc}
Time frame 	& ObsID	& $t_{\rm start}$ [MJD]	& $t_{\rm dur}$ [s] & UVOT \\
\hline
Preflare	& -	& -	& -	& - \\
Night 1		& 00035019176	& 57188.603	& 1996	& U \\
Maximum 	& 00035019180	& 57189.144	& 962	& UVW2 \\
Night 2 	& 00035019181	& 57189.670	& 938	& UVW2 
\end{tabular}
\label{tab:XRT}
\end{table}
The \textit{Neil Gehrels Swift} observatory \citep{Gehrels04} includes three instruments: the Burst Alert Telescope \citep[BAT,][]{Barthelmy05}, the X-ray Telescope \citep[XRT,][]{Burrows05} and the Ultraviolet/Optical Telescope \citep[UVOT,][]{Roming05}. These three instruments provide coverage of the following energy ranges: 5-150\,keV (BAT), 0.3-10\,keV (XRT), and in six optical and ultraviolet filters in the 170-600 nm wavelength range (UVOT). 

XRT data collected in 2015, with Observation Ids 00035019171-00035019188, have been analyzed using version 6.21 of the HEASOFT package.\footnote{\url{http://heasarc.gsfc.nasa.gov/docs/software/lheasoft}} Data calibration has been performed using the \verb|xrtpipeline| procedure and spectral fitting of each single observation has been performed with the \verb|XSPEC| software \citep{Arnaud96}. For the fitting, all observations have been binned to have at least 30 counts per bin and each single observation has been fitted with a single power-law model with a Galactic absorption value of $N_{H} = 2.01 \cdot 10^{20}$\,cm$^{-2}$ \citep{Kalberla2005} set as a frozen parameter. 

The only strictly simultaneous \textit{Swift} observation was during the Maximum time frame. For Night 1 and Night 2, observations have been chosen that were conducted close to the time frames defined in Tab.~\ref{tab:fitres}. The respective Observation IDs, as well as observation times are summarized in Tab.~\ref{tab:XRT}, while the spectral results are given in Tab.~\ref{tab:fitres}. The lightcurve is shown in Fig.~\ref{fig:lc-data}(d) and zoom in on Night 2 in Fig.~\ref{fig:lc-data-zoom}(d).

%
%
\subsection{UV/Optical/IR data}
Simultaneously with XRT, \source\ was monitored in the ultraviolet and optical bands with the UVOT instrument. Observations were taken in six filters: UVW2~(192.8\,nm), UVM2~(224.6\,nm), UVW1~(260.0\,nm), U~(346.5\,nm), B~(439.2\,nm), and V~(546.8\,nm) \citep{Poole2008}. Magnitudes and fluxes have been calculated using \verb|uvotsource| including all photons  from a circular region with radius 5''. In order to determine the background, a circular region with a radius of 10'' located near the source area has been selected. All data points are corrected for dust absorption using the reddening $E(B-V)$ = 0.0245\,mag  \citep{sf11} and the ratios of the extinction to reddening, $A_{\lambda} / E(B-V)$ \citep{Giommi_2006}. Unfortunately, only one UVOT filter was used per Swift pointing (see Tab.~\ref{tab:XRT}) during the flare. Hence, while the resulting fluxes are used in the SED in Fig.~\ref{fig:spec-data}, no lightcurve is shown in Fig.~\ref{fig:lc-data}.

The Automatic Telescope for Optical Monitoring \citep[ATOM,][]{hea04} is a $75\,$cm optical telescope located at the \hess\ site in Namibia. Since 2005, it has monitored around 300 $\gamma$-ray emitters and provides optical data for \hess\ observations.
In 2015, \source\ was monitored with ATOM in the R-band from March until August. Following a rise in flux in June and coinciding with the \hess\ Target-of-Opportunity observations, coverage was increased to up to 20 exposures per night, evenly spread during the time interval from 17h30 to 21h00 UTC. The flux of each observation has been derived using differential photometry using six secondary standard stars from \cite{gea01} in the same field-of-view. The data points have been extinction-corrected similar to the UVOT data. 

SMARTS (Small and Moderate Aperture Research Telescope System) is an optical and infrared telescope dedicated for observations of \fermi\ blazars, visible from the SMARTS site in Chile \citep{Bonning_2012}. 
\source\ has been monitored with the instrument regularly since May 2008. In this paper, the observations collected for the blazar in the season of 2015 in the B, V, R, and J bands have been analyzed. SMARTS data have been corrected for extinction using the corresponding band extinctions from the Galactic Dust Reddening and Extinction Service.\footnote{\url{http://irsa.ipac.caltech.edu/applications/DUST/}}

The R-band lightcurve is shown in Fig.~\ref{fig:lc-data}(e), while the spectral index between the B and J band, calculated as

\begin{align}
    \alpha_{J-B} = \frac{\log{\nu F_J}-\log{\nu F_B}}{\log{\nu_J}-\log{\nu_B}} \label{eq:B-J},
\end{align}
is shown in Fig.~\ref{fig:lc-data}(f). Here, $\nu F_J$ and $\nu F_B$ are the energy fluxes in the J and B band, respectively, while $\nu_J$ and $\nu_B$ are the respective central frequencies of the filters. A zoom-in on the R-band fluxes of Night 2 is shown in Fig.~\ref{fig:lc-data-zoom}(e).

%
%
\subsection{Discussion} \label{sec:mwl_data}
\begin{figure*}[t]
\centering
\includegraphics[width=0.98\textwidth]{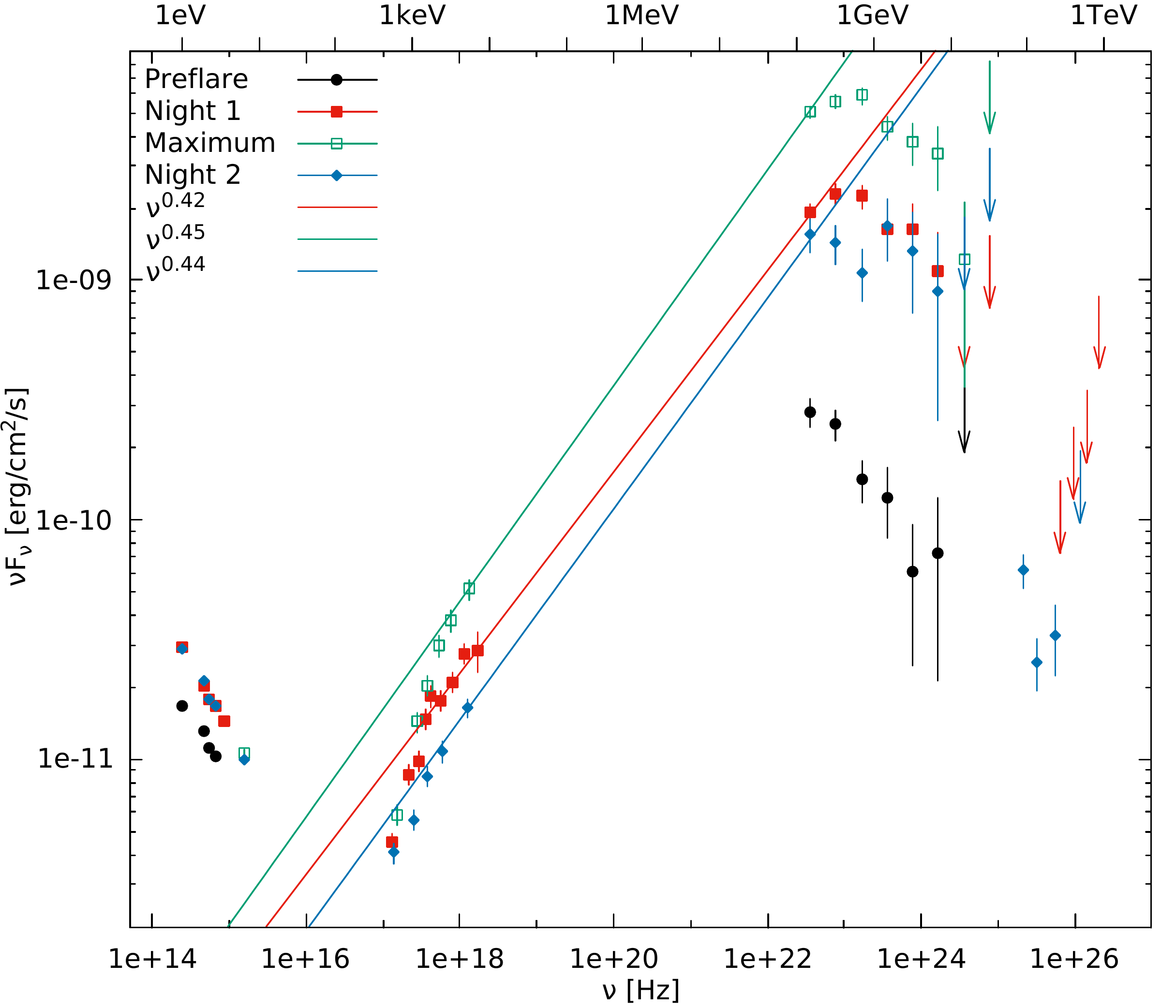}
\caption{Observed multiwavelength SED for the considered time frames with black dots for the Preflare time frame, red filled squares for Night 1, green open squares for the Maximum, and blue diamonds for Night 2. The $\gamma$-ray data have been corrected for EBL absorption using the model by \cite{frv08}. The solid lines show a power-law interpolation for the X-ray to $\gamma$-ray spectrum, as described in the text.}
\label{fig:spec-data}
\end{figure*} 
\begin{table*}[t]
\footnotesize
\caption{Spectral indices of the optical spectrum and interpolation between the X-ray and $\gamma$-ray spectrum. The fourth column gives the energy range of the X-ray to $\gamma$-ray interpolation.}
\begin{tabular}{lccc}
 Time frame 	& $\alpha_{J-B}$		& X-ray--$\gamma$-ray index & $[E_{X},\ E_{\gamma}]$ \\
\hline
Preflare	& $-0.47\pm 0.01$	& - & - \\
Night 1		& $-0.55\pm 0.02$	& $0.42\pm 0.02$ &  $[7.1\,{\rm keV},\ 150\,{\rm MeV}]$  \\
Maximum 	& -			& $0.45\pm 0.01$ &   $[5.5\,{\rm keV},\ 150\,{\rm MeV}]$ \\
Night 2 	& $-0.57\pm 0.01$	& $0.44\pm 0.02$ &   $[5.2\,{\rm keV},\ 150\,{\rm MeV}]$
\end{tabular}
\label{tab:fitres-spec}
\end{table*}
The HE $\gamma$-ray flux, c.f. Fig.~\ref{fig:lc-data}(b), increases by roughly a factor $6$ from the Preflare period to Night 1, followed by another increase by a factor $\sim 3$. The maximum is, hence, a factor $\sim 20$ above the Preflare value. Night 2 is a factor $\sim 4$ below the maximum and about $30\%$ below the Night 1 flux. 

The X-ray flux, c.f. Fig.~\ref{fig:lc-data}(d), increases by a factor $\sim 2$ from Night 1 to the Maximum, and drops subsequently by a factor $\sim 3.5$. These are similar to the ratios of the HE $\gamma$-ray lightcurve and indicate a roughly simultaneous variation of the two bands.

The optical R-band flux rises by about $40\%$ from the Preflare to Night~1, and is at a similar value in Night 2, as is shown in Fig.~\ref{fig:lc-data}(e). The detailed lightcurves from ATOM, as given in Fig.~\ref{fig:lc-data-zoom}(e), indicate minor intranight fluctuations. However, the average value is a good indicator of the optical flux state across the observation window.

Lightcurves are typically exploited to derive a characteristic time scale of a flaring event. For the 2015 flare, \cite{aFea16} derived a flux doubling time scale of less than $5\,$min during the Maximum time frame. However, as the flare bracketed by Nights~1 and 2 lasts for roughly a day, a time scale on the order of minutes is not representative of the whole event. From the HE $\gamma$-ray lightcurve in Fig.~\ref{fig:lc-data}(b), the rise time from the low-point around Night~1 to the Maximum is about $9\,$hrs. The subsequent decay is well described by an exponential function, if the small fluctuations on top of the trend are disregarded. An exponential decay is expected from particle cooling, or if the particles leave the emission region on an energy independent time scale. Performing an exponential fit to the decaying lightcurve, one obtains a time scale of $\sim 9\,$hrs. Hence, this value is considered as the characteristic time scale of the event.

The observed multiwavelength SEDs are shown in Fig.~\ref{fig:spec-data} for the time frames defined in Tab.~\ref{tab:fitres}. In cases where multiple observations are available within a time frame, the data have been averaged. 
The spectral parameters of individual frequency ranges are important for modeling purposes, since they reveal information about the underlying particle distribution. 

The high fluxes during the flaring event allow a precise determination of the spectral index in the HE $\gamma$-ray band in the $3\,$hr time bins, as shown in Fig.~\ref{fig:lc-data}(c). During the flaring event the index is $\sim 2.2$, and hardens significantly to $\sim 2.0$ during the Maximum between Night 1 and 2 (see also Fig.~\ref{fig:lc-data-zoom}(c)). Afterwards the index softens while the flux returns to the quiescence level. At this flux level, the error on the index becomes large for $3\,$hr time bins, and no further conclusions can be drawn as the evolution of the index. The specific parameters for the averaged spectra shown in Fig.~\ref{fig:spec-data} are listed in Tab.~\ref{tab:fitres}.

The X-ray spectrum changes significantly during the flare, as given in Tab.~\ref{tab:fitres}. The spectrum hardens from Night 1 to the Maximum, and softens to Night 2 with the spectrum of Night 2 being even softer than the one in Night 1. Extrapolating the X-ray spectra towards the $\gamma$-ray domain would overpredict the $\gamma$-ray fluxes in all time frames. 

Hence, the broad range of frequencies between the Swift-XRT and \fermi\ spectrum (the explicit energy ranges are given in Tab.~\ref{tab:fitres-spec}) has been interpolated. 
It is assumed that the frequency range can be fitted by a power-law with spectral index $\alpha$, i.e. the energy flux is described by $\nu F_{\nu}\propto \nu^{\alpha}$ with the spectral flux density $F_{\nu}$. The resulting indices are reported in Tab.~\ref{tab:fitres-spec} and the interpolation is shown in Fig.~\ref{fig:spec-data}. The index is positive and constant within errors during the flare with $\alpha\sim 0.44$. Unfortunately, there is no information on the Preflare time frame. 
The indices of the interpolation are softer than the X-ray spectral indices.\footnote{The index of the X-ray ``$\nu F_{\nu}$'' spectrum is $\alpha_{\rm XRT}=2-\Gamma_{\rm XRT}$.} While the X-ray spectra themselves are compatible with simple power-laws, their spectral points and the interpolation lines in Fig.~\ref{fig:spec-data} are suggestive of a break above a few keV.

The indices in the optical energy range between the J and the B band, given in Tab.~\ref{tab:fitres-spec} and shown in Fig.~\ref{fig:lc-data}(f), are derived from the SMARTS observations as described in the previous section. The spectrum softens significantly from the Preflare time frame to the flare, but is roughly constant during Nights~1 and 2. Swift-UVOT observations during the Maximum and Night 2 time frames utilized the UVW2 filter. As can be seen in Fig.~\ref{fig:spec-data}, their fluxes are compatible, and the Night~2 data point agrees well with an extrapolation of the other optical points. This indicates that the optical/UV flux may have been constant during the maximum of the flare. Another possibility could be that the flux in the optical band increased, but the spectrum softened in order to preserve the UV flux.

%
%
\section{The flare in April 2014} \label{sec:2014}
The multiwavelength data of the flare in 2014 were analyzed, modeled and discussed by \cite{pss15} and \cite{Hea15}. 
\cite{pss15} provide a $3\,$hr-binned HE lightcurve obtained with \fermi. This allows one to get the HE $\gamma$-ray fluxes during the \hess\ observation window. They are $\sim 3\E{-6}\,$ph/cm$^2$/s, $\sim 4\E{-6}\,$ph/cm$^2$/s, and $\sim 4\E{-6}\,$ph/cm$^2$/s, respectively. These fluxes coincide with low-points in the lightcurve between separated peaks, similar to Night 1 and Night 2 of the 2015 campaign (c.f. Fig.~\ref{fig:lc-data}). The HE fluxes in 2014 are a factor 2 to 3 lower than during Night 1 and 2 of 2015, which explains the non-detection at VHE energies.

\cite{pss15} produced a HE spectrum integrated over 6 days since MJD~$56749$, which encompasses the \hess\ observations. The average spectrum is significantly curved 
with photon index $\Gamma_{\rm LAT}=2.05\pm 0.05$ and curvature $\beta_{\rm LAT}=0.13\pm 0.03$.\footnote{One should note that a close inspection reveals that the given value for $\beta_{\rm LAT}$ is too small. Better compatibility with the spectral points in Fig.~4 of \cite{pss15} is obtained with $\beta_{\rm LAT}\sim 0.3$.} These parameters are compatible with the parameters obtained in Sec. \ref{sec:fermi} for the Maximum time frame of 2015. The normalization for the \cite{pss15} spectrum is $N_0 = 5.0\E{-6}\,$ph/cm$^2$/s/GeV, about a factor 5 below the normalization of the Maximum time frame in 2015. Extrapolating the \cite{pss15} spectrum to $100\,$GeV (using the corrected value for $\beta_{\rm LAT}$) one obtains an energy flux of $6.7\E{-12}\,$erg/cm$^2$/s, which is below the \hess\ upper limit at that energy (c.f. Fig.~\ref{fig:hess1415}).

\cite{Hea15} derived a HE spectrum for a $6\,$hr time period around the maximum flux (integration time: MJD~$56750.210$--$56750.477$), which is between the first and second night of the \hess\ observations in that year. The derived HE spectrum is compatible with a power-law. The parameters are $\Gamma_{\rm LAT}=2.16\pm 0.06$, and $N_0 = 1.3\E{-5}\,$ph/cm$^2$/s/GeV, which are similar to the parameters obtained for Night 2 in 2015. Hence, a detection at VHE may have been possible during the peak flux in 2014.

\cite{pss15} and \cite{Hea15} used leptonic one-zone models using different combinations of SSC, IC/BLR and IC/DT emission for the high-energy peak. The \hess\ upper limits cannot constrain the models.

%
%
\section{The flare in June 2015} \label{sec:2015}
The significant detection of the \source\ flare with \hess\ in 2015 gives important constraints on the parameter space. These constraints are discussed below, and time-dependent leptonic and lepto-hadronic one-zone models are tested to account for the variability. Most notably, the combined fit of the \fermi\ and H.E.S.S. spectra in Night 2 provides strong constraints on the absorption of $\gamma$-rays, which can be used to constrain the minimum distance of the emission region to the black hole. This is presented first, followed by a brief description of the prevalent thermal photon fields surrounding the jet, which will be used for both modeling attempts. 
%
%
\subsection{Minimum distance of the emission region from the black hole} \label{sec:mindis}
\begin{figure*}[t]
\begin{minipage}{0.49\linewidth}
\centering \resizebox{\hsize}{!}
{\includegraphics{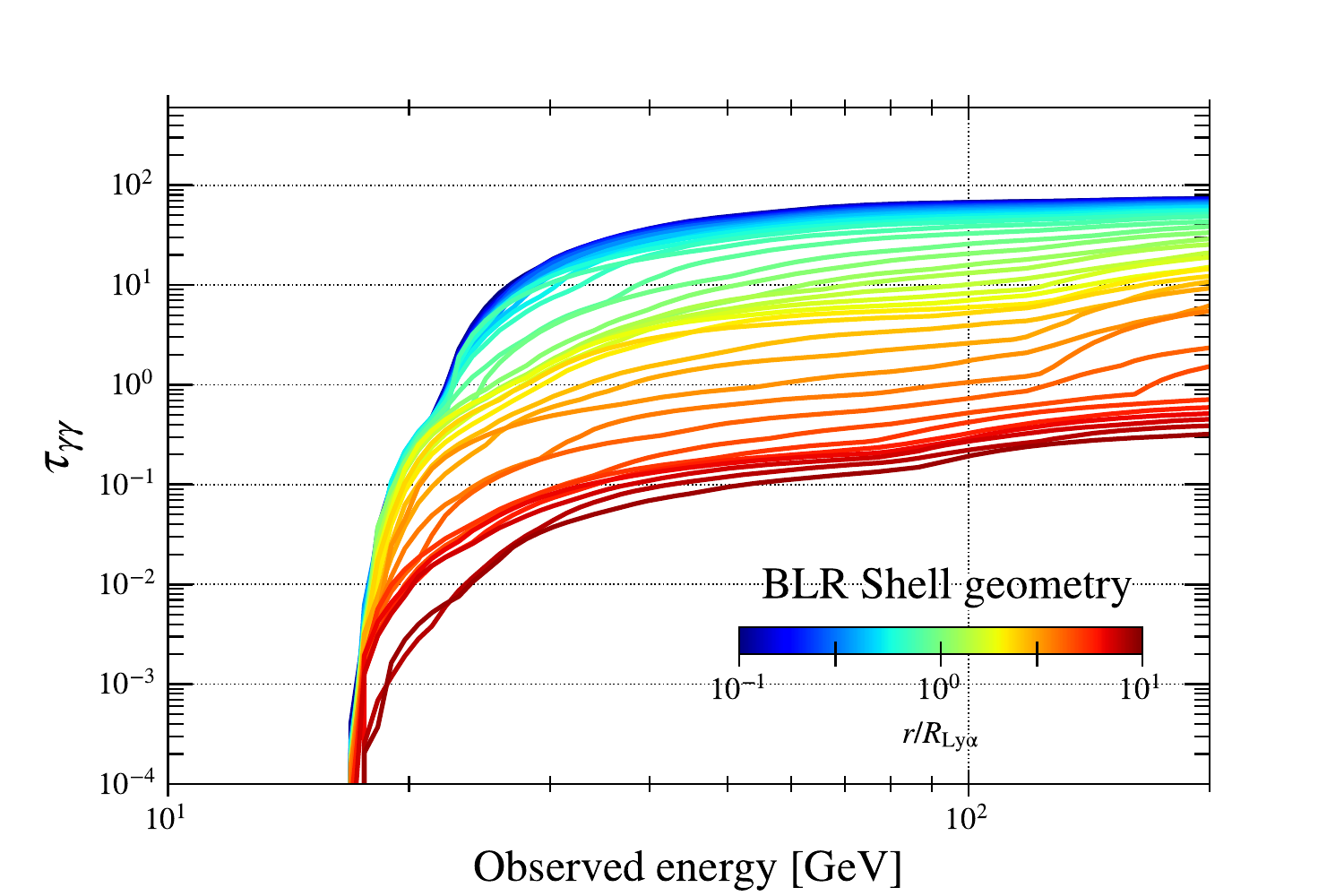}}
\end{minipage}
\hspace{\fill}
\begin{minipage}{0.49\linewidth}
\centering \resizebox{\hsize}{!}
{\includegraphics{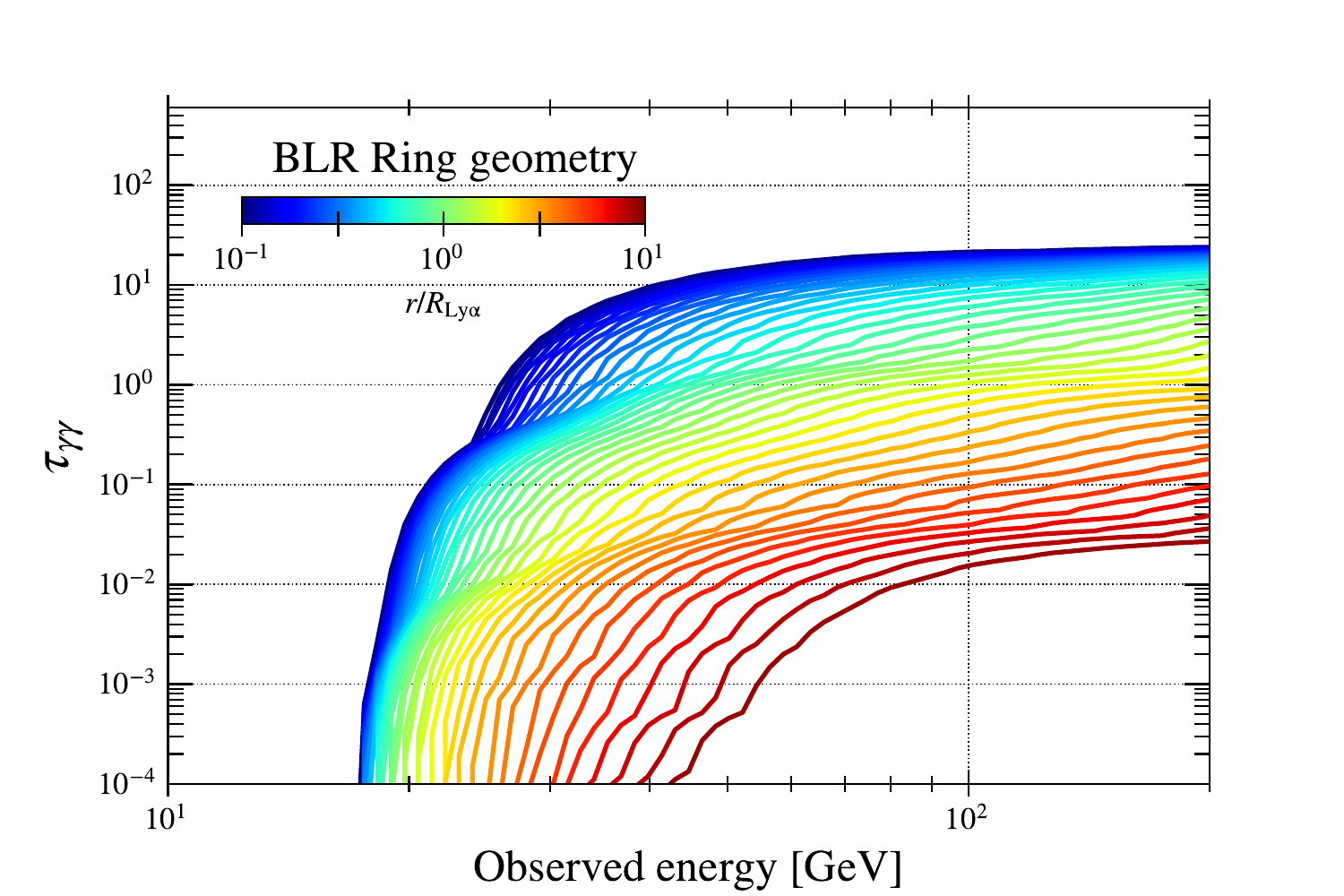}}
\end{minipage}
\caption{Optical depths for $\gamma$-rays emitted along the jet axis at different distances $r$ interacting 
with photons of the BLR emission lines. 
\textit{Left:} BLR modeled with the shell geometry. The crossing of lines at low energies is due to numerical inaccuracies. 
\textit{Right:} BLR modeled with the ring geometry.
The structure in the optical depth are caused by the contributions of different emission lines to the overall optical depth.}
\label{fig:tau}
\end{figure*}
\begin{figure*}[t]
\begin{minipage}{0.49\linewidth}
\centering \resizebox{\hsize}{!}
{\includegraphics{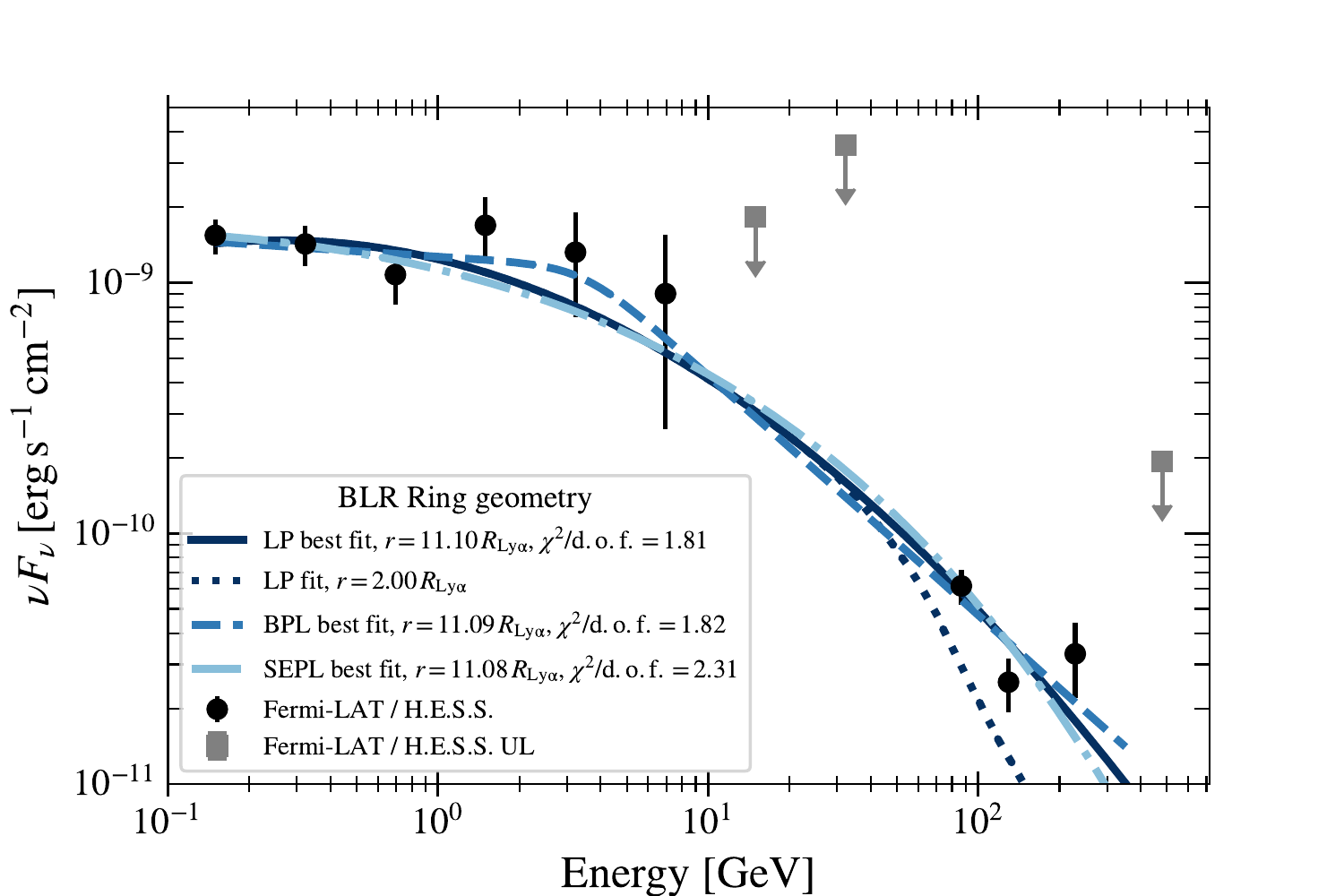}}
\end{minipage}
\hspace{\fill}
\begin{minipage}{0.49\linewidth}
\centering \resizebox{\hsize}{!}
{\includegraphics{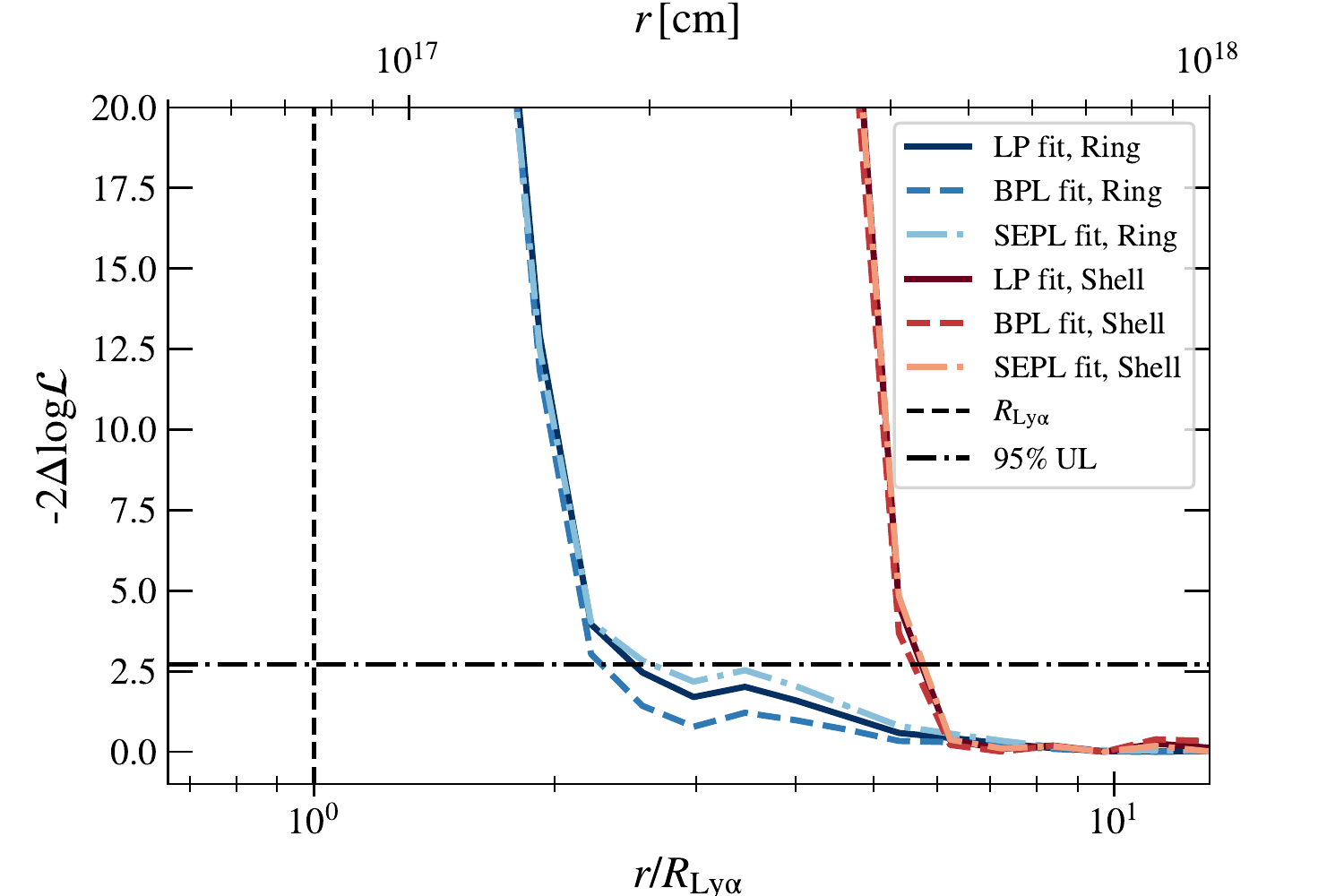}}
\end{minipage}
\caption{\textit{Left:} Best-fit spectra for the BLR ring geometry to the combined \fermi\ and \hess\ data. Both data sets are corrected for EBL absorption following \cite{frv08}.
The spectral shapes do not change significantly, if the shell geometry is assumed instead.
\textit{Right:} Likelihood profile as a function of the distance $r$ for the different assumed intrinsic spectra and BLR geometries.}
\label{fig:fit_blr}
\end{figure*}
The contemporaneous data of \fermi\ and \hess\ enable the search for absorption features caused by pair production of $\gamma$-rays with photons of the BLR. The latter is derived following the model of \cite{Finke:2016}, which is motivated by reverberation mapping and assumes that accretion disc radiation is absorbed by the BLR clouds and re-emitted as monochromatic lines at fixed distances from the black hole. 
The approach here closely follows the method introduced by \cite{Meyer:2019}, who used \fermi\ data of six bright FSRQs to search for absorption features. 

Two geometries of the BLR are implemented in the model. In the \emph{shell} geometry, BLR photons are emitted in infinitesimally thin shells around the black hole, whereas in the \emph{ring} geometry, the BLR photons originate from thin rings orthogonal to the jet axis. The model includes emission lines from Ly$\epsilon$ to H$\alpha$ but neglects any contribution from the thermal continuum. 
Each line has an associated luminosity and is emitted in a shell or a ring at a fixed distance \citep[see Tab.~5 in][]{Finke:2016}. As input the model requires the black hole mass, $M_{\bullet}$, and the luminosity of the H$\beta$ line, $L(\mathrm{H}\beta)$. For \source, $\log_{10} (M_{\bullet} / M_{\odot}) = 8.28$ with the solar mass $M_\odot$, and $L(\mathrm{H}\beta) = 1.7\times10^{43}\,\mathrm{erg}\,\mathrm{s}^{-1}$ are adopted \citep{Liu:2006}. 
Using the relations summarized in \cite{Finke:2016} between $L(\mathrm{H}\beta)$ and $L(5100\,$\AA), as well as between $L(5100\,$\AA) and the radius of the $\mathrm{H}\beta$ emitting shell together with Tab.~5 of \cite{Finke:2016}, the radius of the Ly$\alpha$ emitting shell, $R_{\mathrm{Ly}\alpha} \sim 7.6\times10^{16}\,$cm, is obtained. 
The $\mathrm{Ly}\alpha$ luminosity is the highest in the model (a factor of 12 higher than $L(\mathrm{H}\beta)$) and is therefore responsible for most of the absorption. 
The values for $R_{\mathrm{Ly}\alpha}$ and the Ly$\alpha$ luminosity are broadly consistent with typical values obtained from reverberation mapping~\citep{Kaspi:2007,Bentz:2009,Meyer:2019}
The resulting optical depths, $\tau_{\gamma\gamma}(r,E)$, for both geometries and different distances $r$ of the emission region from the central black hole are shown as a function of the $\gamma$-ray energy in Fig.~\ref{fig:tau}. The shell geometry generally results in higher values of the optical depth \citep[compare also Fig.~14 in][]{Finke:2016}. 
Nevertheless, the optical depths are still lower compared to predictions of more sophisticated BLR models that include continuum emission~\citep[e.g.,][see also the discussion in \citealt{Meyer:2019}]{Abolmasov:2017}. 
In that sense, constraints on the minimum distance between the $\gamma$-ray emitting region and the central black hole can be regarded as conservative. 

The distance $r$ is constrained by simultaneously fitting the \fermi\ and \hess\ data, both corrected for the EBL influence following \cite{frv08}, with an intrinsic spectrum $F(E)$ which is modified by the absorption $\exp(-\tau_{\gamma\gamma})$ (Fig.~\ref{fig:fit_blr}, left). 
The EBL model of \cite{frv08} is in good agreement with other EBL models and with lower limits derived from galaxy number counts \cite[see][for a review]{dk13}.
Since a spectral cut-off due to absorption is degenerate with a cut-off of the intrinsic spectrum, different intrinsic spectral shapes, namely a log-parabola (LP), a power law with sub-exponential cut-off (SEPL) and a broken power law (BPL) are tested. For each combination of intrinsic spectrum and assumed BLR geometry (ring or shell), the parameters of the intrinsic spectrum and $r$ are optimized. This is done using a maximum likelihood optimization, where the likelihood of each \fermi\ and \hess\ spectral flux point is approximated with a Gaussian centered on the measured flux and with a width equal to the flux uncertainty in each bin. One-sided Gaussian distributions are used in case of flux upper limits. 

The resulting best-fit spectra for the ring geometry are shown in the left panel of Fig.~\ref{fig:fit_blr}.
The best-fit values for $r$ are around $\sim 11R_{\mathrm{Ly\alpha}}$ for the ring geometry and around  $\sim 10R_{\mathrm{Ly\alpha}}$ in the shell geometry regardless of tested spectral shapes. The figure includes the $\chi^2$ values per degrees of freedom (d.o.f.). The reduced $\chi^2$ values are all above unity and the fit qualities, measured by the $p$-value of the $\chi^2$ distribution with corresponding d.o.f., are 0.11, 0.12, 0.06 (0.01, 0.01, 0.003) for the LP, BPL, SEPL intrinsic spectra and the ring (shell) geometry, respectively. 
For the LP case, the dotted line additionally shows the case when $r$ is fixed to $2R_{\mathrm{Ly\alpha}}$. For such small values of $r$, the BLR absorption leads to a sharp cut-off of the observed spectrum.
We note that for the SEPL case, a sub-exponential cut-off is preferred by the data. A standard exponential cut-off could reproduce the \fermi\ data and the first two flux points obtained with \hess\ but would under-predict the flux in the highest energy bin by an order of magnitude. 

The right panel of Fig.~\ref{fig:fit_blr} shows the profile likelihood of the fit as a function of $r$. It is evident from the figure that none of the fits significantly prefers the presence of an absorption feature at these large distances over the no-absorption case (which corresponds to the maximum tested distance, $r \sim 30R_{\mathrm{Ly\alpha}}$). Therefore, the maximum likelihood approach is used to derive 95\,\% confidence lower limits on $r$. The lower limits are found by decreasing $r$ until the likelihood increases by $\Delta\ln\mathcal{L} = 2.71 / 2$. All assumed intrinsic spectra result in roughly the same value of the limit of $r \gtrsim 5.4 R_\mathrm{Ly\alpha} = 4.1\times10^{17}\,\mathrm{cm}$. Since the optical depth is smaller for the ring geometry, the lower limit in this case relaxes to $r \gtrsim 2.6 R_\mathrm{Ly\alpha} = 2.0\times10^{17}\,\mathrm{cm}$ for the LP and SEPL intrinsic spectra. The lower limit is slightly lower for the BPL spectrum, $r\gtrsim 2.2R_{\mathrm{Ly\alpha}} = 1.7\times10^{17}\,\mathrm{cm}$.
Note that if only the \emph{Fermi}-LAT data points are fitted with a power law, which is then extrapolated to higher energies including BLR absorption, the flux for all HESS data points is severely under-predicted  for $r \lesssim 7\times 10^{16}\,$cm. This model does not provide a satisfactory fit to the H.E.S.S. data and is especially in tension with the highest energy H.E.S.S. data point, which it under-predicts by more than an order of magnitude. 
In conclusion, the emission zone is confidently placed beyond $r\sim 1.7\times 10^{17}\,$cm (or $3\times 10^3$ Schwarzschild radii), outside the BLR.

%
%
\subsection{The external photon fields} \label{sec:2015gencon}
\begin{table}
\footnotesize
\caption{Parameter description of the external photon fields, symbol and value.}
\begin{tabular}{lcc}
Definition				& Symbol 			& Value \\
\hline
Accretion disk luminosity		& $L_{acc}$			& $3.0\times 10^{45}\,$erg/s \\
BLR luminosity				& $L_{BLR}$			& $2.3\times 10^{44}\,$erg/s \\
BLR radius				& $r_{BLR}$			& $7.6\times 10^{16}\,$cm \\
BLR temperature				& $T_{BLR}$			& $1.0\times 10^{4}\,$K \\
DT luminosity				& $L_{DT}$			& $3.0\times 10^{44}\,$erg/s \\
DT radius				& $r_{DT}$			& $4.2\times 10^{18}\,$cm \\
DT temperature				& $T_{DT}$			& $5.0\times 10^{2}\,$K 
\end{tabular}
\label{tab:inputcom-gen}
\end{table}
In this section, the photon fields external to the jet of \source\ are described. The parameters are listed in Tab.~\ref{tab:inputcom-gen} and are used for the leptonic and lepto-hadronic models described in the next sections. 

The accretion disk is modeled as a Shakura-Sunyaev disk \citep{ss73} with a luminosity $L_{acc}=3.0\times 10^{45}\,$erg/s, which is the average of values given in the literature \citep[e.g.,][]{Hea15,pss15}. The accretion disk luminosity is about $8\%$ of the Eddington power $L_{\rm edd} = 3.78\E{46}\,$erg/s of black hole with mass $M_{\rm bh}\sim 3\times 10^{8}\,M_{\odot}$ \citep[and references therein]{Hea15}. The inner radius of the disk is set to the innermost stable orbit of a Schwarzschild black hole, namely $R_{acc,in}=6\times R_g$ with the gravitational radius of the black hole $R_g$. The outer radius can be estimated following \cite{n15}, and marks the point where the self-gravity of the disk surpasses the gravity of the black hole leading to disk fragmentation further out. For \source\ this corresponds to $R_{acc,out}\sim 430\times R_g$.

Unlike the lines, the thermal BLR parameters are not well known for \source. Using the numbers from the previous section, the radius of the BLR is $r_{BLR}= R_{\mathrm{Ly\alpha}}$, and the luminosity is assumed as $L_{BLR}=2.3\times 10^{44}\,$erg/s. This corresponds to about $8\%$ of the accretion disk luminosity. The given BLR luminosity contains the sum of the line luminosities plus a thermal contribution. The BLR temperature is set to $T_{BLR}=1.0\times 10^{4}\,$K. Note that for the inverse Compton process the BLR line emission can be well approximated by a thermal continuum. 

As the discussion in Sec.~\ref{sec:mindis} indicates that the emission region is located beyond the BLR, its emission may be an inefficient target for the IC process. Whether the strong accretion disk radiation is a useful target field despite being strongly de-boosted, cannot be stated a priori. Therefore, we also invoke the thermal field of a dusty torus, despite the fact that there is no evidence of its presence in  \source. Using estimates from \cite{Hea12}, the radius of the DT becomes $r_{DT} = 4.23\times 10^{18}\,$cm, while the luminosity in this case is assumed to be $10\%$ of the accretion disk. The temperature is assumed to be $T_{DT}=500\,$K.

%
%
\subsection{Leptonic one-zone model} \label{sec:2015lep}
\begin{figure*}[t]
\centering 
\includegraphics[width=0.98\textwidth]{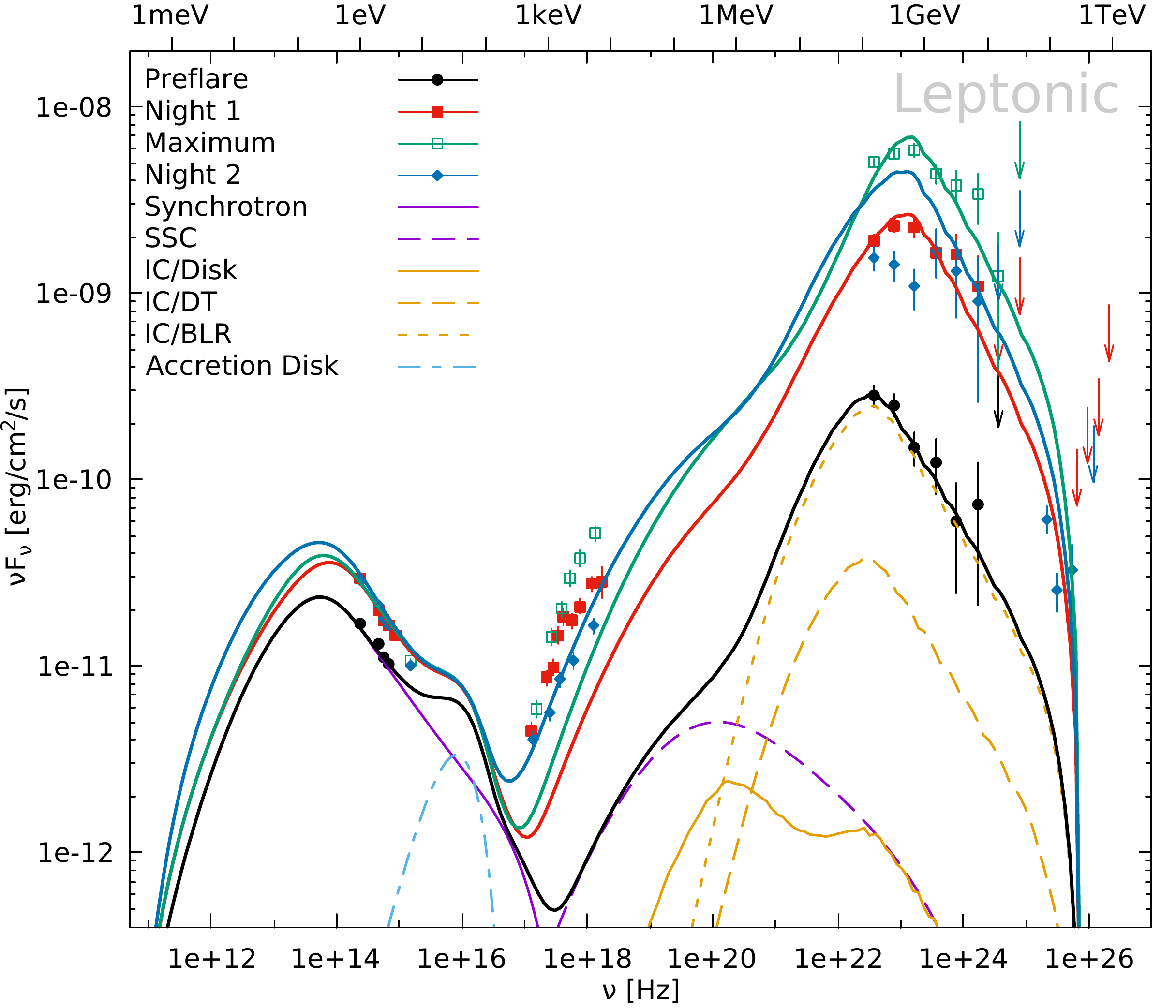}
\caption{Multiwavelength spectra and models for the four time frames: Preflare (black dots), Night 1 (red filled squares), Maximum (green open squares), and Night 2 (blue diamonds). The $\gamma$-ray data have been corrected for EBL absorption using the model by \cite{frv08}. The thick solid lines mark the leptonic models. The thin lines mark spectral components for the Preflare period as labeled. }
\label{fig:spec-lep}
\end{figure*} 
\begin{table}[t]
\footnotesize
\caption{Leptonic model parameter description, symbol and value. Parameters listed below the horizontal line describe the variability.}
\begin{tabular}{lcc}
Definition				& Symbol 			& Value \\
\hline
Emission region distance		& $r\p$				& $1.7\times 10^{17}\,$cm \\ 
Emission region radius			& $R$				& $1.8\times 10^{16}\,$cm \\ 
Doppler factor of emission region	& $\delta$			& $30\,$ \\ 
Magnetic field of emission region	& $B$				& $0.65\,$G \\ 
Electron injection luminosity		& $L_{\rm inj}^e$		& $8.0\times 10^{41}\,$erg/s \\ 
Minimum electron Lorentz factor		& $\gamma_{\rm min}^e$		& $8.0\times 10^2\,$ \\ 
Maximum electron Lorentz factor		& $\gamma_{\rm max}^e$		& $5.0\times 10^4\,$ \\ 
Electron spectral index			& $s^e$				& $2.94\,$ \\ 
Escape time scaling			& $\eta_{\rm esc}$		& $5.0\,$ \\ 
Acceleration to escape time ratio	& $\eta_{\rm acc}$		& $1.0\,$ \\ 
\hline
Magnetic field variation		& $\Delta B_1$			& $-0.39\,$G \\
                		& $\Delta B_2$			& $-0.52\,$G \\
e-injection luminosity variation	& $\Delta L_{{\rm inj},1}^e$	& $6.0\times 10^{42}\,$erg/s \\ 
                        	& $\Delta L_{{\rm inj},2}^e$	& $3.6\times 10^{43}\,$erg/s \\ 
Min. e-Lorentz factor variation		& $\Delta \gamma_{\rm min}^e$	& $8.0\times 10^2\,$ \\ 
e-spectral index variation		& $\Delta s^e$			& $0.18\,$ 
\end{tabular}
\label{tab:inputcom-lep}
\end{table}
The leptonic one-zone model is still the standard model for blazar physics, either in the most fundamental version with synchrotron-self Compton (SSC) or in the slightly extended version with external fields, such as the accretion disk, the broad-line reagion (BLR) and the dusty torus (DT). Its advantage is the relatively low number of parameters, of which a lot can be constrained. From now on parameters marked with a prime are considered in the host galaxy frame, quantities with an asterisk are in the observer's frame, and unmarked quantities are either in the comoving jet frame or invariant. 

The parameters used for the modeling are listed in Tabs.~\ref{tab:inputcom-gen} and \ref{tab:inputcom-lep}. Proper explanations of the parameters and the description of the code are given in App.~\ref{app:lep}. Some of the parameters can be analytically constrained, which is also described in App.~\ref{app:lep}.

The modeling aims to reproduce the flare at the time around the \hess\ observations. Hence, first the Preflare SED is reproduced with the parameters listed above the horizontal line in Tab.~\ref{tab:inputcom-lep}, followed by Night 1. 
Then the Maximum is modeled, after which the evolution is followed to Night 2. The variability is modeled with the following parameter changes: 

\begin{align}
 B (t) &= B + \Delta B_1 \left(\HF{t,\, t\as_{s},\, t\as_{1}} + \HF{t,\, t\as_{m},\, t\as_{2}}\right) \nonumber \\
 &\qquad + \Delta B_2 \HF{t,\, t\as_1,\, t\as_m} \label{eq:dBlep}\\
 L_{\rm inj}^e (t) &= L_{\rm inj}^e + \Delta L_{{\rm inj},1}^e \left(\HF{t,\, t\as_{s},\, t\as_{1}} + \HF{t,\, t\as_{m},\, t\as_{2}}\right) \nonumber \\
 &\qquad + \Delta L_{{\rm inj},2}^e \HF{t,\, t\as_1,\, t\as_m} \label{eq:dLinjelep} \\
 \gamma_{\rm min}^e (t) &= \gamma_{\rm min}^e + \Delta \gamma_{\rm min}^e \HF{t,\, t\as_{s},\, t\as_{2}} \label{eq:dgamminelep} \\
 s^e (t) &= s^e + \Delta s^e \HF{t,\, t\as_{s},\, t\as_{2}} \label{eq:dselep},
\end{align}
where $t\as_s = \mbox{MJD}\, 57186.875$ marks the beginning of the flaring event, $t\as_1 = \mbox{MJD}\, 57188.875$ marks Night 1, $t\as_m = \mbox{MJD}\, 57189.125$ the Maximum and $t\as_2 = \mbox{MJD}\, 57189.875$ Night 2. While these time steps are defined in the observer's frame, they are properly transformed to the comoving frame in the code. The step function $\HF{x,a,b}$ is 1 for $a\leq x\leq b$ and 0 otherwise. Hence, the variability is modeled by 1 or 2 step-function-like changes in the parameters. The variability parameters are listed in Tab.~\ref{tab:inputcom-lep} below the horizontal line. A reasoning for the adopted parameter changes is provided in App.~\ref{app:lep}.

The resulting model SEDs are shown in Fig.~\ref{fig:spec-lep}. The optical regime is dominated by synchrotron photons, while the X-ray regime is mostly SSC, and the \g-ray regime is dominated by the IC/BLR process.
The SEDs are reproduced well for the Preflare, Night 1 and Maximum time frames except in the X-ray domain. However, these time frames can be directly influenced by the changes in the parameters. Subsequently, the injection is returned to Night 1 levels, so the continuing evolution is given by the cooling and escape of the particles. As Night 2 is not reproduced well in the X-ray and $\gamma$-ray energy bands, the chosen parameter set is not adequate to reproduce the decay from the Maximum to Night 2. 

In order to improve the fit in the X-ray domain, a higher SSC flux is required. This could be achieved by a larger number of particles, which would however also increase the synchrotron and IC/BLR fluxes. This could be alleviated by reducing the magnetic field and the luminosity of the BLR. However, the latter is already close to the allowed flux from the line measurements. Increasing the magnetic field, which would in turn increase the SSC flux, would require a brighter BLR in order to preserve the Compton dominance. Additionally, this would require less particles in the emission region. As the SSC flux depends linearly on the magnetic field but quadratically on the particle density, the SSC flux would actually drop. 

The bad fit of Night 2 is driven by the slow particle escape due to the large emission region. Instead of leaving the source, the particles are shifted to lower energies. This has no consequences for the optical domain, which is dominated by particles that cool quickly, explaining the good fit. However, the X-ray and HE $\gamma$-ray domains are dominated by the inverse Compton radiation of less energetic particles. In this energy regime particles have piled up, as the original ones have not yet cooled away, while further particles have reached this energy by cooling down from higher energies. 

This could be alleviated by a faster escape of the particles from the emission region. As the escape is controlled by both the size of the emission region $R$, and the escape time scale parameter $\eta_{\rm esc}$, either of them could be reduced to accelerate the escape. However, $\eta_{\rm esc}$ is already set to only $5.0$, implying that particles remain within the emission region for only five light crossing time scales. This is already a very fast escape, as one expects some diffusion within the emission region due to the magnetic field. 

Hence, reducing $R$ is used to accelerate the escape of particles, as the constraint from the characteristic variability time scale only provides an upper limit. However, reducing $R$ enhances the energy densities of particles and photon fields within the emission region. While this can be accommodated easily for the synchrotron and external-Compton component by reducing the number of injected particles, the SSC flux would drop, as outlined above, and therefore make the fit even worse.

Another possibility is to (additionally) increase the Doppler factor $\delta$. As this value has a direct impact on the internal energy densities of the emission region, the parameters have to be changed considerably. However, also in this case a perfect fit is not possible under the given constraints, which is shown in Fig.~\ref{fig:spec-lep-dop}.

It should be noted that despite the mentioned problems, the \hess\ spectrum is fit well. If the escape problem could be solved, the fit would actually be perfect in the \hess\ domain as the \fermi\ spectra of Night 1 and 2 are similar, and so would be the models. 

As mentioned above, a higher SSC flux could be achieved with a larger number of particles in the jet, while reducing the magnetic field and the external field. While reducing the BLR luminosity is not possible, the emission region could be moved to an even further distance from the black hole, where the DT photon field dominates the external contribution. In fact, parameters can be found that allow for a good fit in large parts of the spectrum, but not perfectly at all energies, c.f. Fig.~\ref{fig:spec-lep-posdop}. The main issue is again the escape of particles, but the delicate interplay of the parameters does not allow to reduce the size of the emission region in this case.

Hence, despite being able to fit the Preflare, Night 1 and Maximum time frames rather well in some cases, the subsequent decay poses a severe problem for the leptonic model. The interplay of the parameters is delicate and requires incredible fine tuning, which could not be achieved for all the details of the spectrum.

\begin{table}[t]
\footnotesize
\caption{Poynting power, proton power, electron power, and radiative power in the observer's frame for the leptonic model curves in Fig.~\ref{fig:spec-lep}. Powers in units of erg/s.}
\begin{tabular}{lcccc}
Symbol 	& Preflare	& Night 1	& Maximum	& Night 2 \\
\hline
$L_B\as$	& $1.8\E{45}$	& $1.9\E{44}$	& $5.3\E{43}$	& $1.9\E{44}$ \\ 
$L_p\as$	& $5.3\E{45}$	& $1.1\E{46}$	& $2.3\E{46}$	& $2.1\E{46}$ \\
$L_e\as$	& $4.9\E{44}$	& $1.8\E{45}$	& $6.9\E{45}$	& $2.7\E{45}$ \\
$L_r\as$	& $2.4\E{45}$	& $1.6\E{46}$	& $3.9\E{46}$	& $3.0\E{46}$ 
\end{tabular}
\label{tab:power-lep}
\end{table}
Nonetheless, it is interesting to study the resulting power output of the model shown in Fig.~\ref{fig:spec-lep}. Tab.~\ref{tab:power-lep} lists the Poynting power, proton power, electron power, and radiative power. The proton power is calculated assuming one cold proton per electron.
The powers have been derived under the assumption that the bulk Lorentz factor is given by $\Gamma_j = \delta$. Compared to the Eddington power of \source's black hole, $L_{\rm edd} = 3.78\E{46}\,$erg/s, the total power is below the Eddinton limit during the Preflare and Night 1 time frames. The Maximum, and Night 2 exceed the Eddington power. By how much depends on the actual value of the mass of the black hole, which has an uncertainty of more than a factor 2 \citep[e.g.,][]{Hea15}. The power output of the jet is dominated by particles and radiation, while the Poynting power is comparable to the other constituents only during the Preflare period. The total power of the jet of Night 2 could be reduced to below the Eddington limit if the emission region contains $90\%$ pairs. Since the radiative output of the jet is already above the Eddington luminosity for the Maximum (keeping the uncertainty in $M_{\rm bh}$ in mind), even a high pair content would not be able to reduce the jet power below that threshold. It should also be noted that the model with a larger Doppler and bulk Lorentz factor (shown in Fig. \ref{fig:spec-lep-dop}) results in super-Eddington jet powers, however with a smaller margin, and a high fraction of pairs may push the total jet power below the Eddington limit in this case.

%
%
\subsection{Lepto-hadronic one-zone model} \label{sec:2015mod}
\begin{figure*}[t]
\centering 
\includegraphics[width=0.98\textwidth]{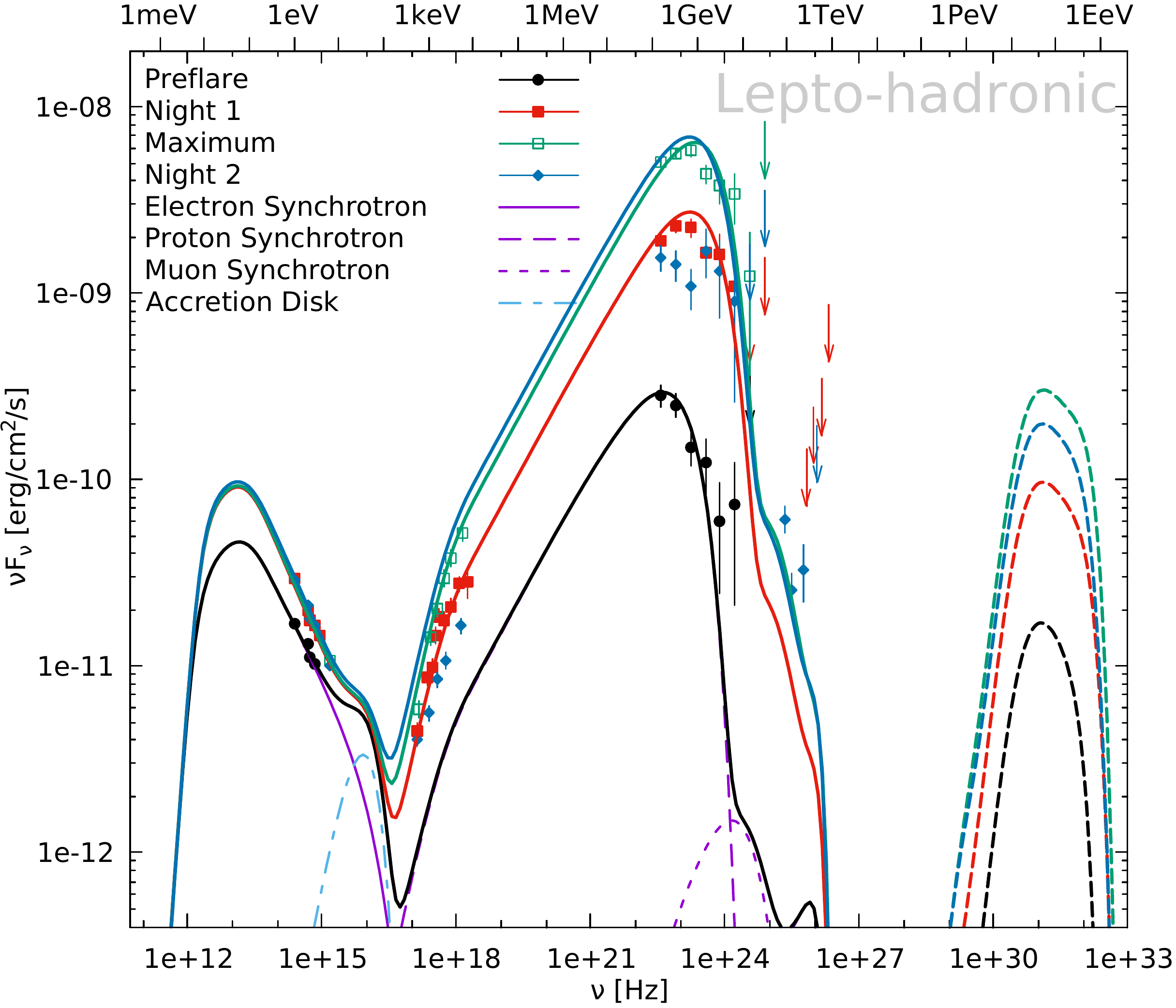}
\caption{Multiwavelength spectra and models for the four considered time frames: Preflare (black dots), Night 1 (red filled squares), Maximum (green open squares), and Night 2 (blue diamonds). The $\gamma$-ray data have been corrected for EBL absorption using the model by \cite{frv08}. The thick solid lines mark the lepto-hadronic photon models, while the thick dashed lines mark the neutrino models. The thin lines mark the spectral components for the Preflare period as labeled. }
\label{fig:spec}
\end{figure*} 
\begin{table}
\footnotesize
\caption{Lepto-hadronic model parameter description, symbol and value. Parameters listed below the horizontal line describe the variability.}
\begin{tabular}{lcc}
Definition				& Symbol 			& Value \\
\hline
Emission region distance		& $r\p$				& $1.7\times 10^{17}\,$cm \\ 
Emission region radius			& $R$				& $1.8\times 10^{16}\,$cm \\ 
Doppler factor of emission region	& $\delta$			& $30\,$ \\ 
Magnetic field of emission region	& $B$				& $50.0\,$G \\ 
Proton injection luminosity		& $L_{\rm inj}^p$		& $7.0\times 10^{43}\,$erg/s \\ 
Minimum proton Lorentz factor		& $\gamma_{\rm min}^p$		& $5.0\times 10^5\,$ \\ 
Maximum proton Lorentz factor		& $\gamma_{\rm max}^p$		& $3.0\times 10^8\,$ \\ 
Proton spectral index			& $s^p$				& $2.11\,$ \\ 
Electron injection luminosity		& $L_{\rm inj}^e$		& $3.3\times 10^{41}\,$erg/s \\ 
Minimum electron Lorentz factor		& $\gamma_{\rm min}^e$		& $5.0\times 10^1\,$ \\ 
Maximum electron Lorentz factor		& $\gamma_{\rm max}^e$		& $2.0\times 10^3\,$ \\ 
Electron spectral index			& $s^e$				& $2.94\,$ \\ 
Escape time scaling			& $\eta_{\rm esc}$		& $5.0\,$ \\ 
Acceleration to escape time ratio	& $\eta_{\rm acc}$		& $30.0\,$ \\ 
\hline
p-injection luminosity variation	& $\Delta L_{{\rm inj},1}^p$	& $4.5\times 10^{44}\,$erg/s \\ 
                            	& $\Delta L_{{\rm inj},2}^p$	& $1.17\times 10^{46}\,$erg/s \\ 
Max. p-Lorentz factor variation		& $\Delta \gamma_{\rm max}^p$	& $3.0\times 10^8\,$ \\ 
e-injection luminosity variation	& $\Delta L_{\rm inj}^e$	& $3.0\times 10^{41}\,$erg/s \\ 
e-spectral index variation		& $\Delta s^e$			& $0.18\,$ 
\end{tabular}
\label{tab:inputcom}
\end{table}
To go beyond the simple one-zone leptonic model, the possibility of a one-zone lepto-hadronic model is explored, following a similar set of source assumptions as made for the leptonic model. Typically, lepto-hadronic models have difficulties in reproducing fast flares owing to the long cooling time scales of protons. However, it was noted by \cite{pea17} that small scale regions with kG magnetic fields could account for the minute-scale variability even in lepto-hadronic models. While the minute-scale variability is not a concern in the present study, it shows the principle possibility to use lepto-hadronic models to account for the 2015 flare of \source.

The parameters reproducing the Preflare period are listed in Tabs.~\ref{tab:inputcom-gen} and \ref{tab:inputcom} above the horizontal line. They are explained along with a discussion of the constraints and the details of the code in App.~\ref{app:had}. 

Again, a self-consistent reproduction of the \source\ spectra is attempted by changing input parameters as follows:

\begin{align}
 L_{\rm inj}^p (t) &= L_{\rm inj}^p + \Delta L_{{\rm inj},1}^p \HF{t,\, t\as_{s},\, t\as_1} + \Delta L_{{\rm inj},2}^p \HF{t,\, t\as_1,\, t\as_1} \label{eq:dLinjp1} \\
 \gamma_{\rm max}^p (t) &= \gamma_{\rm max}^p + \Delta \gamma_{\rm max}^p \HF{t,\, t\as_{s},\, t\as_{2}} \label{eq:dgammaxp1} \\
 L_{\rm inj}^e (t) &= L_{\rm inj}^e + \Delta L_{\rm inj}^e \HF{t,\, t\as_{s},\, t\as_{2}} \label{eq:dLinjehad} \\
 s^e (t) &= s^e + \Delta s^e \HF{t,\, t\as_{s},\, t\as_{2}} \label{eq:dsehad},
\end{align}
where the time steps are the same as in the leptonic case. The maximum proton Lorentz factor, the electron injection luminosity, and the electron spectral index are only varied once during the flare and remain at their levels until the end of the flare. The proton injection, however, is varied until the beginning of the Maximum, with a single injection on top of that, after which it returns to Preflare levels. The variability parameters are listed in Tab.~\ref{tab:inputcom} below the horizontal line. The magnetic field is not varied, as there is no constraint on it in this case.

The four derived spectra for the lepto-hadronic model are shown in Fig.~\ref{fig:spec}. The optical component is well reproduced by electron synchrotron emission in all cases. The X-ray and HE \g-rays are dominated by proton synchrotron emission, while the VHE \g-rays are influenced by muon synchrotron emission with some contributions from synchrotron emission of secondary electrons. The Preflare HE $\gamma$-ray spectrum is well matched except for the highest energy bins. Night 1 is well reproduced for both X-rays and HE $\gamma$-rays. The Maximum is well reproduced in the HE $\gamma$-rays, and X-rays. Night 2 is significantly overpredicted in both the X-rays and HE $\gamma$-rays, while the VHE $\gamma$-rays are too low.

The main problem is, again, the slow escape of particles, which in this case is mainly the proton escape, since the electrons cool very efficiently in the strong magnetic field. However, the protons barely cool nor escape, which is why the hadronic spectral components of Night 2 even slightly exceed those of the Maximum. As before, reducing the size of the emission region $R$ would increase particle escape, and hence reduce the flux. The increase in internal energy densities would increase the production of pions, muons and secondary electron/positron pairs. This could produce a flux that is compatible with the VHE spectrum. However, tests have revealed that the interplay of escape and cooling -- while weak -- has an observable effect, which makes a fit in either the X-rays or the VHE $\gamma$-rays problematic.

As discussed below, the jet power significantly exceeds the Eddington power of the black hole. This could be mitigated by increasing the bulk Lorentz and Doppler factor, as the same radiative output requires less power in the particles. A realization is shown in Fig.~\ref{fig:spec-had-dop}. However, while the total jet power decreases slightly, it still significantly surpasses the Eddington power. One should also note that despite a significantly smaller source size, and a much faster escape it is still not possible to fit the X-ray and VHE $\gamma$-ray spectrum of Night 2 simultaneously, as the latter is underpredicted by the model. As the emission region is placed at the minimum distance allowed by the result of Sec.~\ref{sec:mindis}, the external fields cannot be enhanced further to allow for a larger number of muons to be produced, as their synchrotron emission is responsible for the VHE $\gamma$-ray output within this model.

It should be noted that the spectral characteristics in the X-ray domain require a rather large minimum proton Lorentz factor. This is difficult to explain through conventional acceleration processes, which expect a minimum proton Lorentz factor of $\sim 1$.

Hence, despite being less constrained than the leptonic model, the one-zone lepto-hadronic model is also not able to self-consistently reproduce the observed characteristics of the flare. 

\begin{table}[t]
\footnotesize
\caption{Poynting power, proton power, electron/positron power, and radiative power in the observer's frame for the lepto-hadronic model curves in Fig.~\ref{fig:spec}. Powers in units of erg/s.}
\begin{tabular}{lcccc}
Symbol 	& Preflare	& Night 1	& Maximum	& Night 2 \\
\hline
$L_B\as$	& $2.7\E{48}$	& $2.7\E{48}$	& $2.7\E{48}$	& $2.7\E{48}$ \\ 
$L_p\as$	& $1.9\E{47}$	& $1.1\E{48}$	& $3.3\E{48}$	& $2.2\E{48}$ \\
$L_e\as$	& $5.2\E{42}$	& $1.1\E{43}$	& $1.2\E{43}$	& $1.3\E{43}$ \\
$L_r\as$	& $2.3\E{45}$	& $1.9\E{46}$	& $4.2\E{46}$	& $4.6\E{46}$ 
\end{tabular}
\label{tab:power-had}
\end{table}
The Poynting, proton, electron/positron and radiative powers are given in Tab.~\ref{tab:power-had}. The power output in this case is dominated by the Poynting flux and the proton power, while the radiative and electron/positron powers are subdominant. The electron/positron power increases throughout the flare and even during Night 2. This comes from the ongoing injection of secondary electron/positron pairs from the muon decay, which inject highly energetic pairs that carry a large amount of power. In all cases the total power significantly exceeds the Eddington power, $L_{\rm edd} = 3.78\E{46}\,$erg/s. The general picture does not change much for a larger bulk Lorentz factor.

The decay of pions and muons releases neutrinos, and the model neutrino spectra arriving at Earth are shown in Fig.~\ref{fig:spec} for the four specific time steps. 
Using IceCube's effective area \citep{icc13}, the detectable neutrino rates for IceCube can be calculated, and hence the potential number of detectable neutrinos from the event. 
Concentrating on the $\sim 27\,$h time window bounded by the \hess\ observations and which covers the peak flux in the HE band, the number of detectable neutrinos is $5\times 10^{-4}$. 
Even if the emission region would be located within the BLR, which would result in a larger pion and muon production rate and hence a larger number of neutrinos, the rate would not increase enough in order to reach unity \citep{zea19}. Hence, no neutrino is expected to be detected by IceCube during the Maximum of the flare, and this approach cannot be used to distinguish between the leptonic and lepto-hadronic models.

%
%
\subsection{Discussion} \label{sec:2015dis}
The most important result is the lower limit of the distance of the emission region from the black hole, placing it outside the BLR.
This directly implies that the observed minute-scale variability in the HE $\gamma$-ray band \citep{aFea16} is not caused by an emission region encompassing the entire width of the jet. It rather points towards small emission regions or turbulent cells within a larger active region \citep[e.g.,][]{g13,m14}. Furthermore, it adds to the growing evidence \citep[e.g.,][]{zea17,cea18} that jets can produce $\gamma$-ray emission on large distances from the black hole. 

None of the one-zone models can fully reproduce the observed characteristics of the 2015 flare in \source, and the jet powers are a severe constraint for the models. The leptonic model is mostly below the Eddington power of the black hole. However, it surpasses the Eddington power during the Maximum time frame. Interestingly, during this time frame the radiative power emitted by \source\ is already very close to or even surpasses the Eddington limit depending on the actual mass of the black hole. This underlines the extreme nature of this flare. In the lepto-hadronic model, the Eddington power is surpassed during every time frame by a large factor. This is a common problem of proton-synchrotron models \citep[e.g.,][]{zb15}. 
This might be possible for a short flare, as the one described here, but it is unlikely for a longer period, such as the Preflare time frame, which resembles the ground state, where the total power of the model is $\sim 3\times 10^{48}\,$erg/s. The assumption that the jet power is provided dominantly by the accretion power, implies a radiative efficiency of the accretion disk of less than $10^{-3}$, using the bolometric disk luminosity given Tab.~\ref{tab:inputcom-gen}. This is much less than the typical radiative efficiency of accretion disks of $0.1$-$0.2$ in active galaxies, and unlikely on long time scales \citep[see][for a detailed discussion]{zb15}.
Hence, while the flare itself might be hadronically induced, the quiescent state is probably not.

In this work, it is assumed that the emission region is a standing shock, like a recollimation shock, within the jet and does not change its position during the flare. The bulk flow is, thus, provided by the jet material crossing the shock. However, a moving shock would cover a distance of $\sim 1\,$pc during the flare. In such a scenario, the external fields would change with time, which could explain the reduction in \g-ray flux at the end of the flare without a faster escape of particles.
Additionally, more sophisticated models, such as a spine-in-sheath model \citep{gtc05}, a jets-in-jet model \citep{g13}, a moving mirror model \citep{vtc17}, and others, might provide an improved description of the observations. However, testing these possibilities is beyond the scope of this paper.

The failure of the simple leptonic one-zone model in \source\ has been noted before \citep[e.g.,][]{brm09,bea16}. 
The 2015 flare has been explicitly modeled by \cite{bea16} using stationary leptonic and hadronic one-zone models. The discussion in \cite{bea16} focuses on \textit{INTEGRAL} observations conducted for $\sim 14\,$h around the peak of the $\gamma$-ray outburst. All data were integrated over this time bin, including the significant variability in all bands. \cite{bea16} conclude that their leptonic model would not produce VHE $\gamma$-ray emission. On the other hand, their hadronic model would allow for VHE $\gamma$-ray emission. 
This strong statement cannot be confirmed here, as the time-dependent leptonic and lepto-hadronic models allow for VHE $\gamma$-ray emission, even though a self-consistent fit could not be achieved. 

%
%
\section{Limits on Lorentz invariance violation} \label{sec:liv}
\begin{figure*}[t]
\begin{minipage}{0.49\linewidth}
\centering \resizebox{\hsize}{!}
{\includegraphics{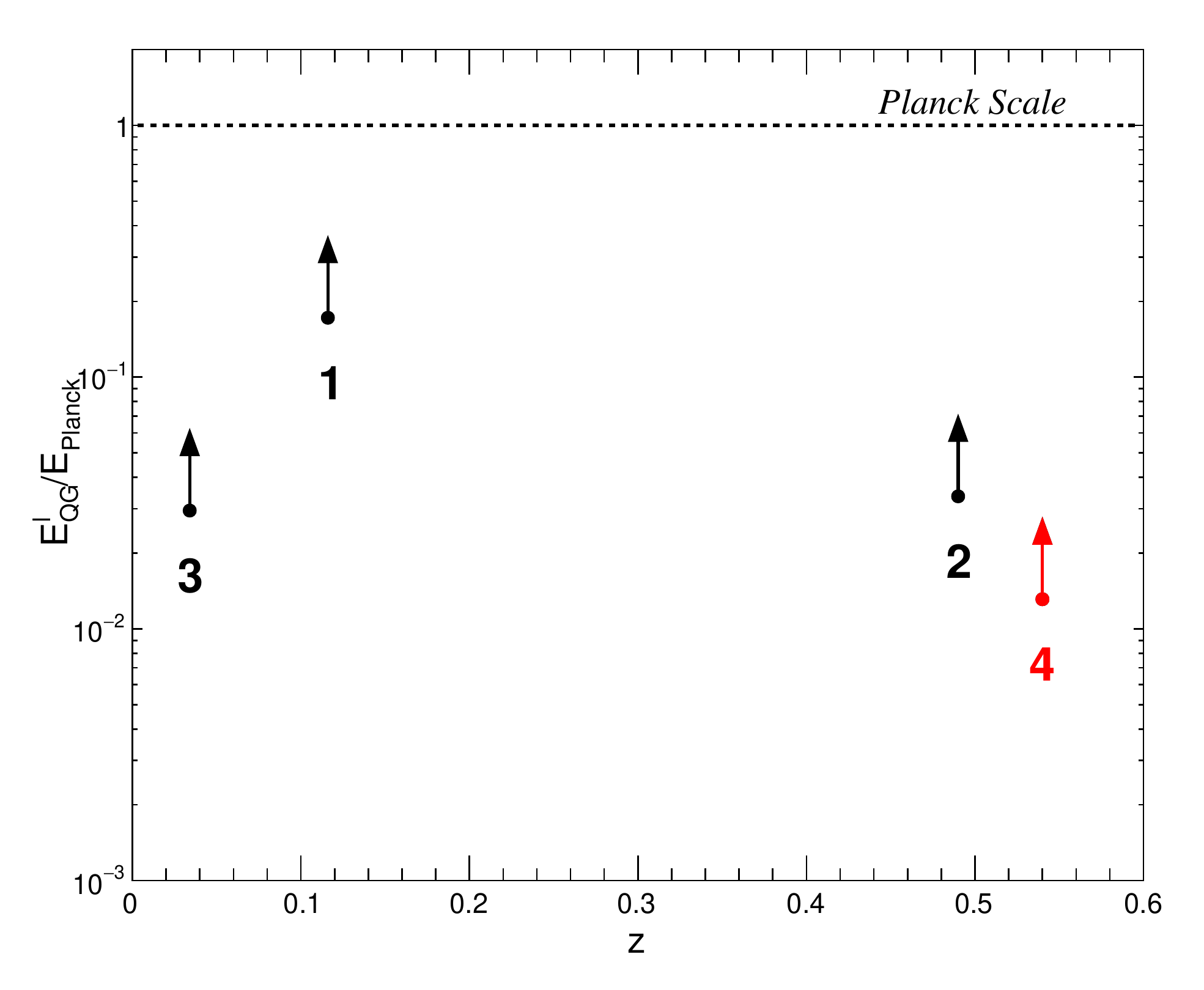}}
\end{minipage}
\hspace{\fill}
\begin{minipage}{0.49\linewidth}
\centering \resizebox{\hsize}{!}
{\includegraphics{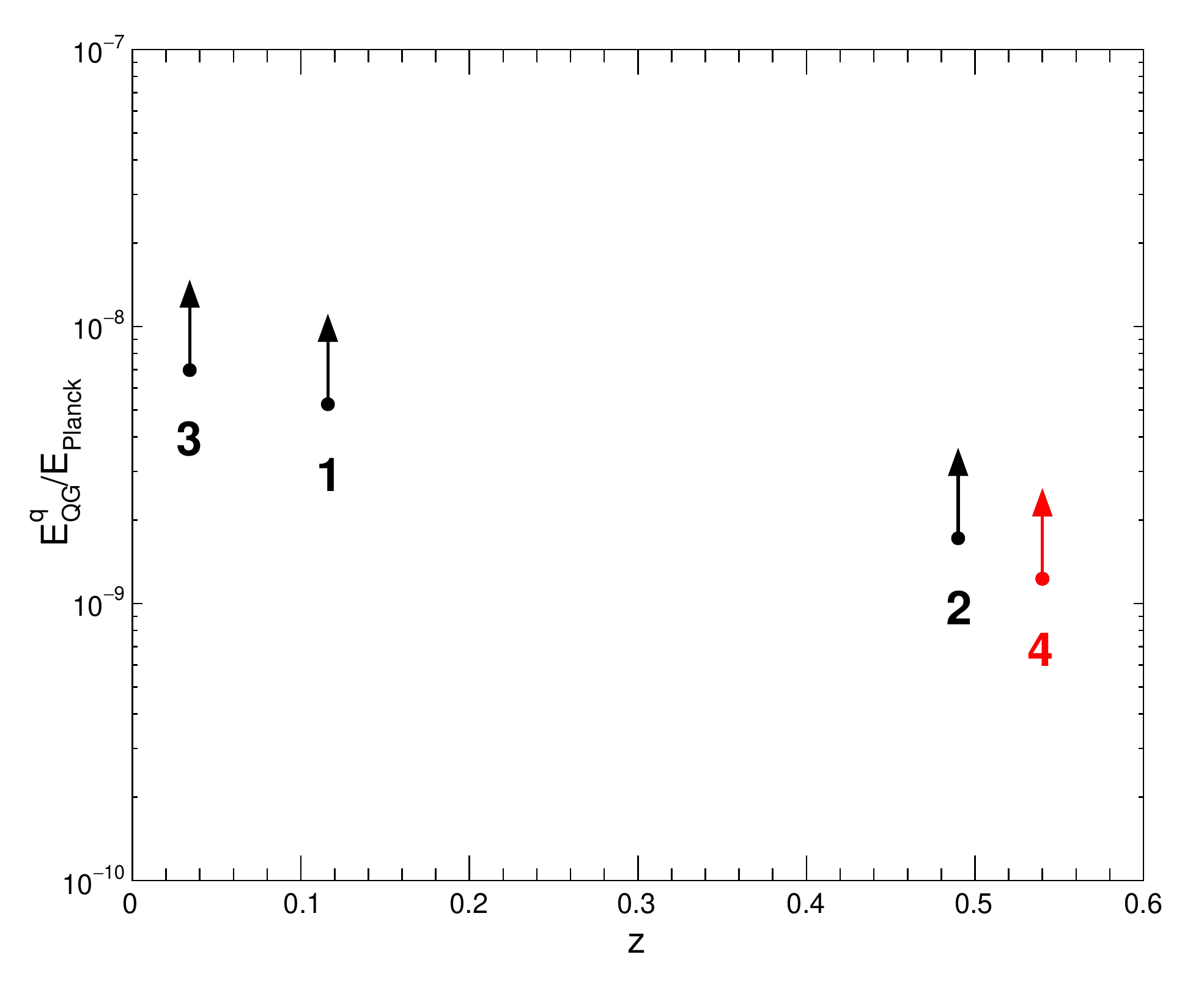}}
\end{minipage}
\caption{Lower limits on the sub-luminal linear (\textit{left}) and quadratic (\textit{right}) terms of the modified dispersion relation obtained with \hess\ for several AGN as a function of redshift. 1: PKS 2155-304 \citep{aHea11}, 2: PG 1553+113 \citep{aHea15}, 3: Mrk 501 \citep{cea15}, 4: \source\ (this work).}
\label{fig:liv}
\end{figure*}
\begin{table}
\footnotesize
\caption{The 95\% limits on $E^l_\mathrm{QG}$ and $E^q_\mathrm{QG}$ derived from the \source\ observations by \hess\ in the sub-luminal and super-luminal cases.}
\begin{tabular}{lcc}
					& Sub-luminal	& Super-luminal \\
\hline
Linear effect $E^l_\mathrm{QG}$		&$> 1.6\times 10^{17}\,$GeV	& $> 3.5\times 10^{17}\,$GeV \\
Quadratic effect $E^q_\mathrm{QG}$	&$> 1.5\times 10^{10}\,$GeV	& $> 1.8\times 10^{10}\,$GeV 

\end{tabular}
\label{tab:liv}
\end{table}
Several models of Quantum Gravity (QG) predict a violation of Lorentz Invariance (LIV in the following for \textit{Lorentz Invariance Violation}) in the form of a modified dispersion relation for photons in vacuum \citep[see][for a general review about QG phenomenology, including modified dispersion relations]{amelino13}. This effect should be dominant at energy scales of the order of the Planck scale ($\sim10^{19}$~GeV) but it is believed that it could be observed at lower energies, though linearly or quadratically suppressed.
The modified dispersion relation leads to an energy-dependent speed of light. High energy photons propagating in vacuum may either be slower (the sub-luminal case) or faster (the super-luminal case) than low energy photons. 
In addition, the longer the propagation distance is, the larger the time delay between photons of different energies should be.

Variable or transient astrophysical sources are then very good candidates to constrain $E^l_\mathrm{QG}$ and $E^q_\mathrm{QG}$, respectively the energy scales for linear and quadratic LIV effects. 
Stringent limits have been obtained with flaring AGN \citep{aHea11} and pulsars \citep{Ahnen17}. The best limits available so far \citep{vea13}, obtained with Gamma-Ray Bursts \object{GRB 090510} and \object{GRB 090926A}, are above the Planck scale for $E^l_\mathrm{QG}$ but still need to be confirmed with more objects.
\source\ is a prime candidate to perform such a study, owing to its high redshift and high number of VHE $\gamma$-ray photons detected during the 2015 flare (Night 2).

The search for energy-dependent time-delays was performed with the likelihood method first introduced by \cite{me09}, and adapted to take the background contribution into account as detailed by \cite{aHea15} for the case of PG~1553+113 flare of 2012. The likelihood method ``compares'' the arrival times of photons at high energies (here in the 300~GeV - 2~TeV range, corresponding to 185~events) with a parameterized template lightcurve obtained at low energies (100~GeV - 150~GeV, 243~events). As in \cite{aHea11,aHea15}, a detailed study was performed using simulations to evaluate statistical and systematic errors. Statistical errors were calibrated by generating 300 realizations of the lightcurve mimicking the data, with no lag. As in the case of PG~1553+113 flare of 2012 which shows similar characteristics of variability and background contamination, systematic errors were found to be mainly due to the low energy template lightcurve parameterization and to energy selections. Overall, no significant lag was measured and one-sided 95\% confidence level limits on $E^l_\mathrm{QG}$ and $E^q_\mathrm{QG}$ were computed. These results are given in Tab.~\ref{tab:liv} for the sub- and super-luminal cases. Sub-luminal limits are also shown in Fig.~\ref{fig:liv} together with other results published by the \hess\ Collaboration \citep{aHea11,aHea15,cea15}.

Even if less constraining than those obtained from other studies, the \source\ flare results will be valuable for future population studies due to the high redshift of this source. 

%
%
\section{Summary \& Conclusions} \label{sec:sumcon}
The FSRQ \source\ underwent two major HE $\gamma$-ray outbursts in April 2014 and June 2015. These were among the brightest flares detected with \fermi, and during the 2015 flare a HE $\gamma$-ray variability time scale of $\sim 5\,$min was detected \citep{aFea16}. Both flares were followed up with the VHE $\gamma$-ray experiment \hess. 
The observations in 2014 have not yielded a detection,
and the upper limits are not particularly constraining for modeling attempts of that event.
However, a significant detection of \source\ at VHE $\gamma$-ray has been achieved during the 2015 event with $8.7\,\sigma$ above an energy threshold of $66\,$GeV. This allows one to derive strong constraints on source parameters.

Most importantly, the VHE $\gamma$-ray spectrum along with a simultaneous HE spectrum can be used to derive the amount of absorption of the emitted $\gamma$-rays through the BLR photon field. This can be translated into an estimate of the distance of the emission region from the black hole. An elaborate analysis \citep{Meyer:2019} results in a lower limit (95\% confidence level) on the distance at $r\gtrsim 1.7\times10^{17}\,$cm, placing the emission region outside of the BLR. 

In this work, using both a time-dependent leptonic, and a time-dependent lepto-hadronic one-zone model, a reproduction of the contemporaneous spectra up to VHE $\gamma$-ray energies and the hour-scale variability has been attempted. 
The leptonic model reproduces the data well in the optical and \g-ray bands for most time frames except for Night 2, where the spectrum is overproduced in the \g-ray component. The X-ray spectrum cannot be adequately fit. Accelerating the escape of particles after the Maximum by invoking a smaller emission region along with other parameter changes, does not improve the fit significantly. 

The lepto-hadronic model faces similar difficulties, as the decrease from the Maximum to Night 2 in the high-energy component is also not well covered. Changing the parameters is also unable to provide a satisfactory fit.
The number of model neutrinos is too low to be detectable by IceCube, and can therefore not be used as a discriminator of the models.

In summary, simple one-zone models cannot fully reproduce the observed characteristics of the 2015 flare in \source\ within the given constraints, and more elaborate models are required. 

The lower limits on LIV linear and quadratic energy scales obtained in this study are comparable to those derived from other flaring AGNs observed with similar characteristics of variability and background level. The data described here will be included in the overall combination of LIV study results which is currently being prepared by the three major IACT experiments \citep[see][for a preliminar study from simulated data]{nea17}. Due to its high redshift, \source\ will also be an interesting target for population studies with the future Cherenkov Telescope Array, which is expected to greatly improve the current limits on QG energy scale.

%
%
\begin{acknowledgement}

The support of the Namibian authorities and of the University of Namibia in facilitating 
the construction and operation of H.E.S.S. is gratefully acknowledged, as is the support 
by the German Ministry for Education and Research (BMBF), the Max Planck Society, the 
German Research Foundation (DFG), the Helmholtz Association, the Alexander von Humboldt Foundation, 
the French Ministry of Higher Education, Research and Innovation, the Centre National de la 
Recherche Scientifique (CNRS/IN2P3 and CNRS/INSU), the Commissariat à l’énergie atomique 
et aux énergies alternatives (CEA), the U.K. Science and Technology Facilities Council (STFC), 
the Knut and Alice Wallenberg Foundation, the National Science Centre, Poland grant no. 2016/22/M/ST9/00382, 
the South African Department of Science and Technology and National Research Foundation, the 
University of Namibia, the National Commission on Research, Science \& Technology of Namibia (NCRST), 
the Austrian Federal Ministry of Education, Science and Research and the Austrian Science Fund (FWF), 
the Australian Research Council (ARC), the Japan Society for the Promotion of Science and by the 
University of Amsterdam. We appreciate the excellent work of the technical support staff in Berlin, 
Zeuthen, Heidelberg, Palaiseau, Paris, Saclay, Tübingen and in Namibia in the construction and 
operation of the equipment. This work benefited from services provided by the H.E.S.S. 
Virtual Organisation, supported by the national resource providers of the EGI Federation. \\
A.W. is supported by Polish National Agency for Academic Exchange (NAWA).
This paper has made use of up-to-date SMARTS optical/near-infrared lightcurves that are available at \url{www.astro.yale.edu/smarts/glast/home.php}. This research has made use of the NASA/IPAC Infrared Science Archive, which is operated by the Jet Propulsion Laboratory, California Institute of Technology, under contract with the National Aeronautics and Space Administration.

\end{acknowledgement}
%
%

%

%
%
\begin{appendix}
\section{Leptonic code and model constraints} \label{app:lep}
The time-dependent leptonic code used in this work was developed by \cite{db14}. The calculations are performed in the comoving frame of the emission region, and the calculated spectra and lightcurves are subsequently transformed into the observer's frame with the Doppler factor $\delta$ and the redshift $\zr$. At each time-step a power-law distribution of electrons with injection luminosity $L_{\rm inj}^{e}$, minimum and maximum Lorentz factor $\gamma_{\rm min}^e$ and $\gamma_{\rm max}^e$, and spectral index $s^e$ is injected into the spherical emission region of radius $R$, which is pervaded by a tangled magnetic field $B$. The particle distribution is then evolved self-consistently following a Fokker-Planck equation, considering stochastic acceleration, radiative cooling, and catastrophic losses. The Fokker-Planck equation is solved using a Crank-Nicholson scheme. The stochastic acceleration time scale is parameterized as a multiple $\eta_{\rm acc}$ of the escape time scale, which itself is a multiple $\eta_{\rm esc}$ of the light-crossing time scale $R/c$. Any of these time scales is independent of energy implying a ``hard-sphere'' magnetic field turbulence model for the acceleration term. It should be noted that the acceleration in this case merely acts as a mild re-acceleration of particles. The main acceleration is induced by the injection spectrum, which could originate from a small acceleration region \citep[as in the models of, e.g.,][]{ws15,cpb15} that is not accounted for here. The electrons are subject to synchrotron and inverse-Compton cooling including SSC, IC/Disk, IC/BLR, and IC/DT. The latter three depend on the distance $r$ of the emission region from the black hole. The final particle distribution at the end of each time-step is considered in the next time-step with new particles injected on top, and the cycle repeats. It should be noted that the emitted radiation is also self-consistently calculated following the radiative transport equation. This implies that not all emitted photons leave the emission region instantaneously in each time step. Some remain behind and are used in the next time step for all mentioned processes. Eventually an equilibrium solution is found for the particles, where injection, acceleration, and losses balance. Subsequently, any parameter may be disturbed for one or more time steps in order to produce an outburst, after which the code follows the particle evolution until the original equilibrium solution is reached again. 

The code has been slightly expanded to include the absorption of $\gamma$-rays while they traverse the external photon fields. This adopts the methodology of \cite{be16}. Additionally, the BLR and DT routines have been slightly expanded to allow for anisotropic photon distributions.

Several constraints can be inferred from the data. The high observed luminosity of the flare along with the short variability implies a large Doppler factor $\delta$ in order to keep the particle energy densities low. Unfortunately, no direct constraint on the value of the Doppler factor is available. However, observations of moving radio knots revealed apparent speeds of up to $\sim 21c$ \citep{lea13} in the radio jet of \source, also implying large Doppler factors. For the (main) modeling, $\delta=30$ is adopted, which is well within bounds of the observed apparent superluminal motion \citep[see also][]{Hea15}.

The standard constraint on the size of the emission region is by equating the characteristic flare time scale with the light-crossing time scale of the emission region. Using the value of the characteristic flare time scale from Sec.~\ref{sec:mwl_data}, the radius $R$ becomes

\begin{align}
 R \leq \frac{\delta t_{\rm char}c}{1+\zr} = 1.8\times 10^{16} \est{\delta}{30}{} \,\mbox{cm} \label{eq:Rconst}.
\end{align}
This is the maximum value allowed by the characteristic time scale. 

The spectral index of the electron distribution is directly related to the spectral index of the synchrotron component. In the strong cooling regime, the spectral index, $\alpha$, and the electron spectral index $s^e$ are related by $s^e = 2-2\alpha$, where the electron distribution is $n^e(\gamma)\propto \gamma^{-s^e}$. It has been verified a posteriori that the cooling is indeed in the strong cooling domain. Using the spectral index values for the IR to UV regime from Tab.~\ref{tab:fitres-spec}, the electron spectral index during the Preflare period is $s^e = 2.94\pm 0.01$, while during the flare it is $s^e = 3.12 \pm 0.03$. For the latter, the average value of Night 1 and Night 2 has been used, since they are consistent within errors. The electron indices are softer than expected by conventional acceleration processes. However, they are in line with typical electron indices derived for \source\ \citep{brm09} and other FSRQs \citep[e.g.,][]{vea11,bea14,zea19}.

Simple considerations of the IC process, especially with thermal target photons, lead to the estimate that in a restricted energy range the resulting IC component depends similarly on the electron spectral index as in the synchrotron component. Hence, the spectral index in the $\gamma$-ray domain probed by \fermi, $\alpha = 2-\Gamma_{\rm LAT}$, should be comparable to the spectral index in the IR to UV domain. Tabs.~\ref{tab:fitres} and \ref{tab:fitres-spec} indicate that for the Preflare period the indices are similar, while during the flare the hardening in the $\gamma$-ray domain does not correspond to the softening in the IR to UV regime. This can be mitigated by increasing the minimum electron Lorentz factor $\gamma_{\rm min}^{e}$ during the flare.

In order to model the variability, four parameters have been varied as given in Eqs.~(\ref{eq:dBlep}) to (\ref{eq:dselep}). The changes are inspired by either direct measurements, as in the case of the spectral index, or inferences from the changes in the spectrum, such as the Compton dominance. The latter implies a change in the ratio from the external photon density to the magnetic field density. It is assumed that the external thermal fields do not change during the short flare. Hence, the magnetic field must decrease to account for an increase in the Compton dominance. As the \fermi\ spectra are almost identical in Night 1 and 2, the Compton dominance is the same in these two nights, which is why the same magnetic field strength is required in both nights (giving the reason for, in total, three Heaviside functions in Eq.~(\ref{eq:dBlep})). The requirement to increase $\gamma_{\rm min}^{e}$ has been mentioned before. The flux changes in the synchrotron component imply an increase in particle energy density in order to compensate the decrease in the magnetic field.

\begin{figure}[t]
\centering 
\includegraphics[width=0.48\textwidth]{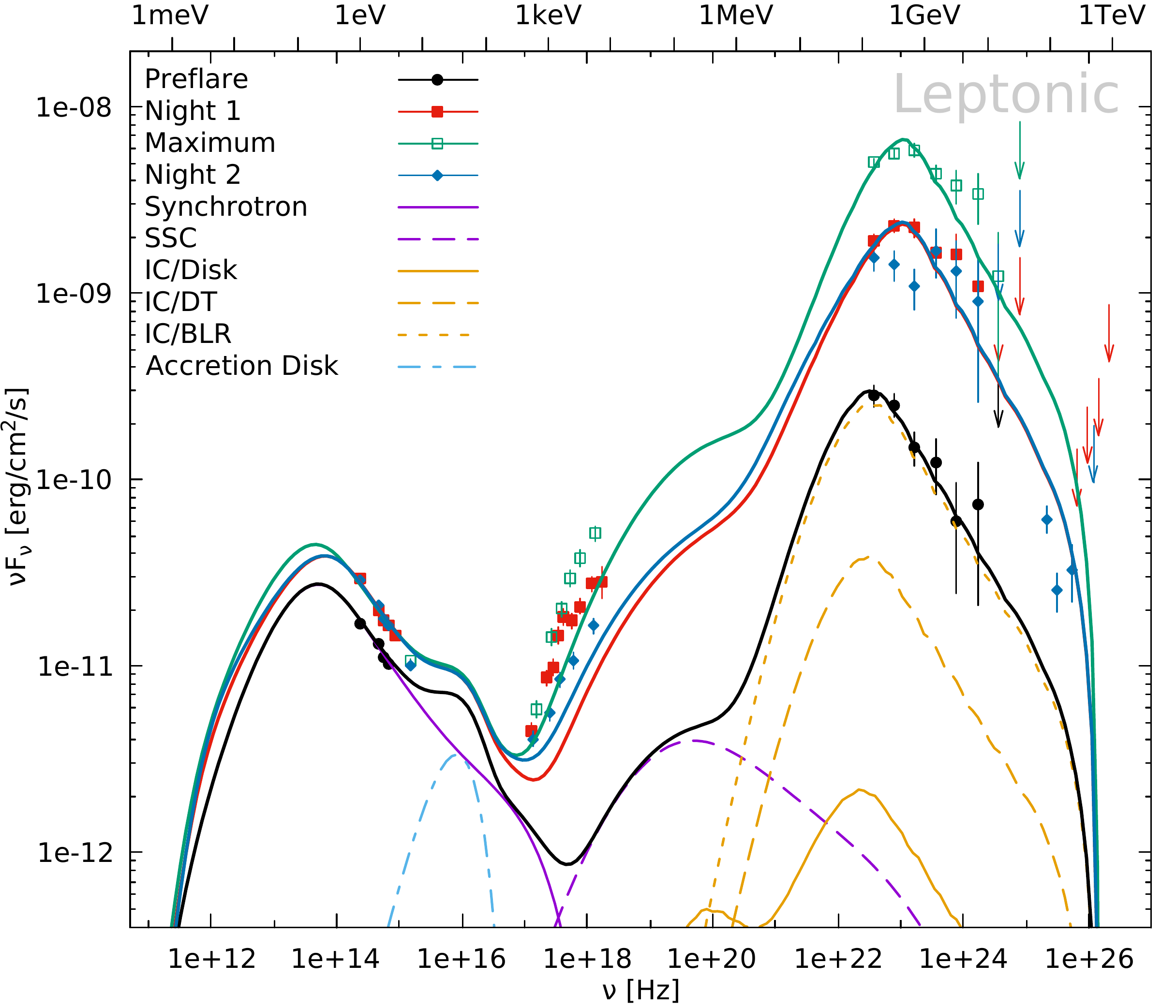}
\caption{Leptonic model using a larger Doppler factor along with the parameters in Tab.~\ref{tab:inputcom-lep-dop}. Data and model lines as in Fig.~\ref{fig:spec-lep}.}
\label{fig:spec-lep-dop}
\end{figure}

\begin{figure}[t]
\centering 
\includegraphics[width=0.48\textwidth]{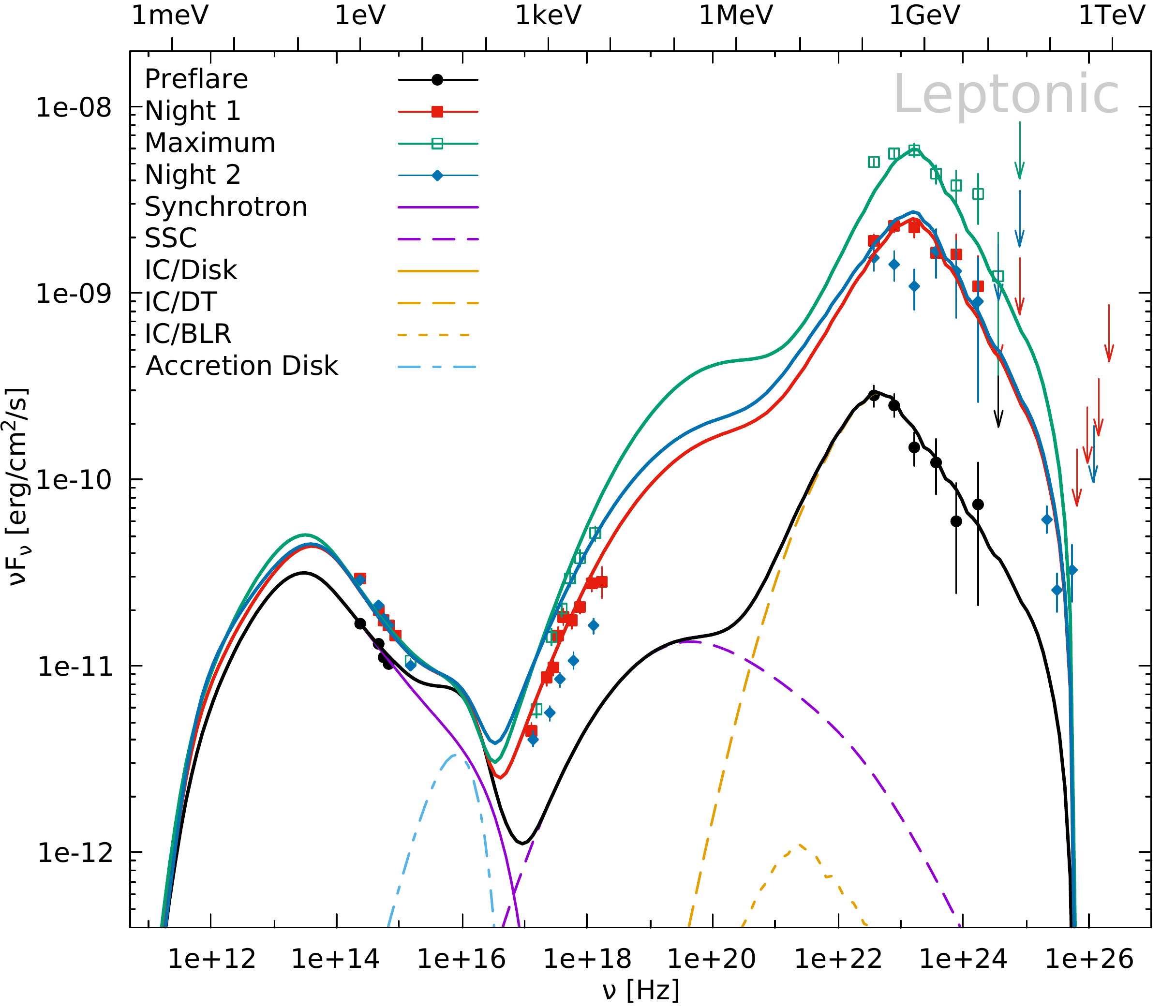}
\caption{Leptonic model using a larger distance from the black hole and a larger Doppler factor. The parameters are given in Tab.~\ref{tab:inputcom-lep-posdop}. Data and model lines as in Fig.~\ref{fig:spec-lep}.}
\label{fig:spec-lep-posdop}
\end{figure}

Figs.~\ref{fig:spec-lep-dop} and \ref{fig:spec-lep-posdop} show additional leptonic models with a larger Doppler factor, and a larger distance from the black hole, respectively. The respective parameter sets are given in Tabs.~\ref{tab:inputcom-lep-dop} and \ref{tab:inputcom-lep-posdop}. The variability follows the same dependencies as given in Eqs.~(\ref{eq:dBlep}) to (\ref{eq:dselep}).

\begin{table}[t]
\footnotesize
\caption{Leptonic model with larger Doppler factor, Fig.~\ref{fig:spec-lep-dop}. Parameter description, symbol and value. Parameters listed below the horizontal line describe the variability.}
\begin{tabular}{lcc}
Definition				& Symbol 			& Value \\
\hline
Emission region distance		& $r\p$				& $1.7\times 10^{17}\,$cm \\ 
Emission region radius			& $R$				& $6.0\times 10^{15}\,$cm \\ 
Doppler factor of emission region	& $\delta$			& $50\,$ \\ 
Magnetic field of emission region	& $B$				& $0.90\,$G \\ 
Electron injection luminosity		& $L_{\rm inj}^e$		& $1.5\times 10^{41}\,$erg/s \\ 
Minimum electron Lorentz factor		& $\gamma_{\rm min}^e$		& $5.0\times 10^2\,$ \\ 
Maximum electron Lorentz factor		& $\gamma_{\rm max}^e$		& $4.0\times 10^5\,$ \\ 
Electron spectral index			& $s^e$				& $2.94\,$ \\ 
Escape time scaling			& $\eta_{\rm esc}$		& $5.0\,$ \\ 
Acceleration to escape time ratio	& $\eta_{\rm acc}$		& $1.0\,$ \\ 
\hline
Magnetic field variation		& $\Delta B_1$			& $-0.52\,$G \\
                		& $\Delta B_2$			& $-0.67\,$G \\
e-injection luminosity variation	& $\Delta L_{{\rm inj},1}^e$	& $1.2\times 10^{42}\,$erg/s \\ 
                        	& $\Delta L_{{\rm inj},2}^e$	& $3.36\times 10^{42}\,$erg/s \\ 
Min. e-Lorentz factor variation		& $\Delta \gamma_{\rm min}^e$	& $4.0\times 10^2\,$ \\ 
e-spectral index variation		& $\Delta s^e$			& $0.18\,$ 
\end{tabular}
\label{tab:inputcom-lep-dop}
\end{table}
\begin{table}[t]
\footnotesize
\caption{Leptonic model with larger distance from the black hole and Doppler factor, Fig.~\ref{fig:spec-lep-posdop}. Parameter description, symbol and value. Parameters listed below the horizontal line describe the variability.}
\begin{tabular}{lcc}
Definition				& Symbol 			& Value \\
\hline
Emission region distance		& $r\p$				& $1.0\times 10^{18}\,$cm \\ 
Emission region radius			& $R$				& $1.0\times 10^{16}\,$cm \\ 
Doppler factor of emission region	& $\delta$			& $50\,$ \\ 
Magnetic field of emission region	& $B$				& $0.35\,$G \\ 
Electron injection luminosity		& $L_{\rm inj}^e$		& $8.0\times 10^{41}\,$erg/s \\ 
Minimum electron Lorentz factor		& $\gamma_{\rm min}^e$		& $6.0\times 10^2\,$ \\ 
Maximum electron Lorentz factor		& $\gamma_{\rm max}^e$		& $3.0\times 10^4\,$ \\ 
Electron spectral index			& $s^e$				& $2.94\,$ \\ 
Escape time scaling			& $\eta_{\rm esc}$		& $5.0\,$ \\ 
Acceleration to escape time ratio	& $\eta_{\rm acc}$		& $1.0\,$ \\ 
\hline
Magnetic field variation		& $\Delta B_1$			& $-0.21\,$G \\
                		& $\Delta B_2$			& $-0.26\,$G \\
e-injection luminosity variation	& $\Delta L_{{\rm inj},1}^e$	& $6.0\times 10^{42}\,$erg/s \\ 
                        	& $\Delta L_{{\rm inj},2}^e$	& $2.1\times 10^{43}\,$erg/s \\ 
Min. e-Lorentz factor variation		& $\Delta \gamma_{\rm min}^e$	& $6.0\times 10^2\,$ \\ 
e-spectral index variation		& $\Delta s^e$			& $0.18\,$ 
\end{tabular}
\label{tab:inputcom-lep-posdop}
\end{table}
%

%
\section{Lepto-hadronic code and model constraints} \label{app:had}
The time-dependent lepto-hadronic code used in this work was developed by \cite{db16} and extended to include external photon fields in \cite{zea19}. This includes the same possibilities as in the leptonic code. Namely, the absorption of \g-rays in the external fields and anisotropic external fields. The code works similarly to the leptonic code described above with the addition of the proton distribution and related effects. In addition to the electrons, protons are injected at each time step with a power-law distribution with injection luminosity $L_{\rm inj}^{p}$, minimum and maximum Lorentz factor $\gamma_{\rm min}^p$ and $\gamma_{\rm max}^p$, and spectral index $s^p$. The protons follow the Fokker-Planck equation with the same structure as the electrons. However, next to synchrotron cooling, protons can also interact with ambient photon fields to produce pions. While the neutral pions are assumed to instantaneously decay into $\gamma$-rays, the charged pions decay into muons, which subsequently decay into electrons or positrons. The evolution of the charged pions and muons is also calculated by a Fokker-Planck equation, considering the same effects as for the protons and electrons. The electrons and positrons from the muon decay are used as an additional injection term for the electron evolution. All charged particles are subject to radiative cooling, which is considered to be synchrotron emission. For electrons Compton losses on the ambient fields are also considered. It turns out that these are subdominant. The neutrino spectra produced during pion and muon decay are also calculated. The time-dependency of the code is achieved as in the leptonic case through variations of a few parameters. 

Below, the constraints for the lepto-hadronic model are described. Several of the leptonic constraints are reused, most notably the Doppler factor and the size of the emission region.

The spectral indices for the particle distributions can be derived from the observed spectral indices of the observed spectrum listed in Tab.~\ref{tab:fitres-spec}. In fact, for the electrons the result is unchanged. From the interpolated X-ray to $\gamma$-ray spectrum one can deduce the proton spectral index assuming slow cooling of the protons. This has also been verified a posteriori. The relation between the observed spectral index $\alpha$ and the proton spectral index $s^p$ is $s^p=3-2\alpha$, where the proton distribution is described by $n^p(\gamma) = \gamma^{-s^p}$. Using the values of $\alpha$ listed in Tab.~\ref{tab:fitres-spec}, the proton spectral index for Night 1 is $2.16\pm 0.04$, $2.10\pm 0.02$ for the Maximum, and $2.11\pm 0.04$ for Night 2. These are compatible within errors, and are kept constant during the modeling.

The hardening of the HE $\gamma$-ray spectrum is mimicked by increasing the maximum proton Lorentz factor during the flare. The apparent break in the X-ray domain is accounted for by a large minimum proton Lorentz factor.

\begin{figure}[t]
\centering 
\includegraphics[width=0.48\textwidth]{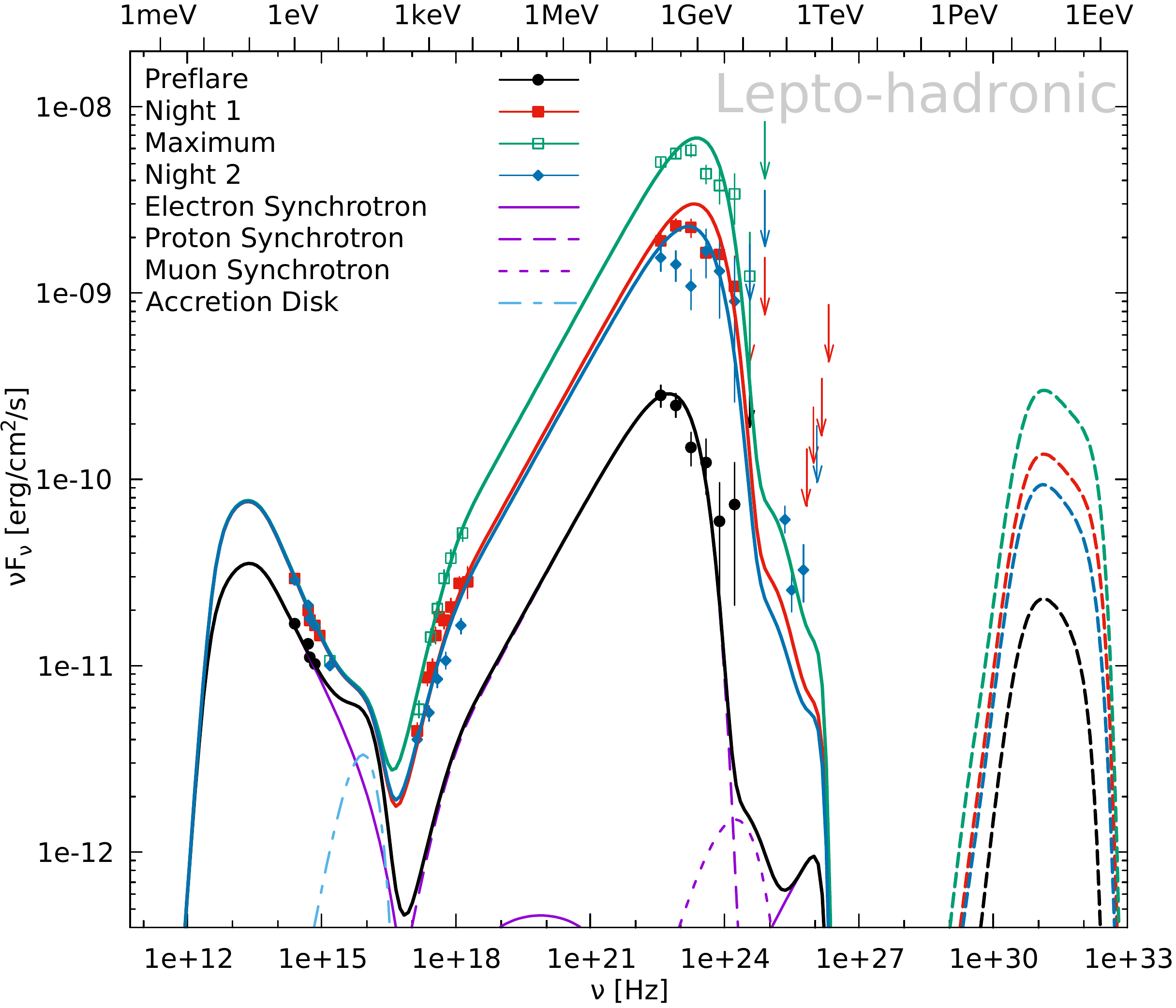}
\caption{Lepto-hadronic model using a larger Doppler factor along with the parameters in Tab.~\ref{tab:inputcom-had-dop}. Data and model lines as in Fig.~\ref{fig:spec}.}
\label{fig:spec-had-dop}
\end{figure}

Fig.~\ref{fig:spec-had-dop}
shows an additional lepto-hadronic model with a larger Doppler factor. The parameter set is given in Tab.~\ref{tab:inputcom-had-dop}. 
The variability follows the same dependencies as given in Eqs.~(\ref{eq:dLinjp1}) to (\ref{eq:dsehad}).

\begin{table}[t]
\footnotesize
\caption{Lepto-hadronic model with larger Doppler factor, Fig.~\ref{fig:spec-had-dop}. Parameter description, symbol and value. Parameters listed below the horizontal line describe the variability.}
\begin{tabular}{lcc}
Definition				& Symbol 			& Value \\
\hline
Emission region distance		& $r\p$				& $1.7\times 10^{17}\,$cm \\ 
Emission region radius			& $R$				& $4.5\times 10^{15}\,$cm \\ 
Doppler factor of emission region	& $\delta$			& $50\,$ \\ 
Magnetic field of emission region	& $B$				& $50.0\,$G \\ 
Proton injection luminosity		& $L_{\rm inj}^p$		& $3.0\times 10^{43}\,$erg/s \\ 
Minimum proton Lorentz factor		& $\gamma_{\rm min}^p$		& $4.0\times 10^5\,$ \\ 
Maximum proton Lorentz factor		& $\gamma_{\rm max}^p$		& $2.5\times 10^8\,$ \\ 
Proton spectral index			& $s^p$				& $2.11\,$ \\ 
Electron injection luminosity		& $L_{\rm inj}^e$		& $3.3\times 10^{40}\,$erg/s \\ 
Minimum electron Lorentz factor		& $\gamma_{\rm min}^e$		& $5.0\times 10^1\,$ \\ 
Maximum electron Lorentz factor		& $\gamma_{\rm max}^e$		& $2.0\times 10^3\,$ \\ 
Electron spectral index			& $s^e$				& $2.94\,$ \\ 
Escape time scaling			& $\eta_{\rm esc}$		& $5.0\,$ \\ 
Acceleration to escape time ratio	& $\eta_{\rm acc}$		& $30.0\,$ \\ 
\hline
p-injection luminosity variation	& $\Delta L_{{\rm inj},1}^p$	& $1.6\times 10^{44}\,$erg/s \\ 
                            	& $\Delta L_{{\rm inj},2}^p$	& $1.6\times 10^{45}\,$erg/s \\ 
Max. p-Lorentz factor variation		& $\Delta \gamma_{\rm max}^p$	& $2.0\times 10^8\,$ \\ 
e-injection luminosity variation	& $\Delta L_{\rm inj}^e$	& $3.5\times 10^{40}\,$erg/s \\ 
e-spectral index variation		& $\Delta s^e$			& $0.18\,$ 
\end{tabular}
\label{tab:inputcom-had-dop}
\end{table}
\end{appendix}
\end{document}